\documentclass[12pt,preprint]{aastex}
\usepackage{epstopdf}
\usepackage{graphicx,rotating}
\usepackage{amssymb}
\usepackage{natbib}
\newcommand{\beq}{\begin{equation}}
\newcommand{\eeq}{\end{equation}}
\newcommand{\etal}{{\sl et~al.~}}
\newcommand{\kms}{km s$^{-1}$}
\newcommand{\kp}{{$\kappa$} Pav}
\newcommand{\HST}{{\it HST}}
\newcommand{\HIPs}{{\sl Hipparcos~}}

\newcommand{\HIP}{{\sl Hipparcos}}

\def\fdg{\hbox{$.\!\!^\circ$}}
\def\arcmin{\hbox{$^\prime$}}
\def\XZpi{$ 1.67\pm0.17$~}
\def\SUpi{$ 1.42\pm0.16$~}
\def\RZpi{$ 2.12\pm0.16$~}
\def\RRpi{$ 3.77\pm0.13$~}
\def\UVpi{$ 1.71\pm0.10$~}
\def\VYpi{$ 6.44\pm0.23$~}
\def\kppi{$ 5.57\pm0.28$~}
\begin{document}

\received{}
\revised{}
\accepted{}

\shorttitle{Population II Distance Scale Zero-Points}
\shortauthors{Benedict \etal}

\bibliographystyle{/Active/my2}
\title{Distance Scale Zero-Points from Galactic RR Lyrae Star
Parallaxes\footnote{Based on observations made with the NASA/ESA Hubble Space Telescope, obtained at the Space Telescope Science Institute, which is operated by the Association of Universities for Research in Astronomy, Inc., under NASA contract NAS5-26555} }

\author{ G.\ Fritz  Benedict\altaffilmark{2}, Barbara E.
McArthur\altaffilmark{2}, Michael W. Feast\altaffilmark{3}$^,$\altaffilmark{9}, Thomas G. Barnes\altaffilmark{2}, Thomas E. Harrison\altaffilmark{4}$^,$\altaffilmark{8}, Jacob L. Bean\altaffilmark{12}$^,$\altaffilmark{17}, John W. Menzies\altaffilmark{9}, Brian Chaboyer\altaffilmark{16}, Luca Fossati\altaffilmark{13}, Nicole Nesvacil\altaffilmark{10}$^,$\altaffilmark{19}, Horace A. Smith\altaffilmark{7}, Katrien Kolenberg\altaffilmark{10}$^,$\altaffilmark{12}, C. D. Laney\altaffilmark{9}$^,$\altaffilmark{15}, Oleg Kochukhov\altaffilmark{14}, Edmund P. Nelan\altaffilmark{11}, D.~V. Shulyak\altaffilmark{18}, Denise Taylor\altaffilmark{11}, Wendy L. Freedman\altaffilmark{6}}

\altaffiltext{2}{McDonald Observatory, University of Texas, Austin, TX 78712}
\altaffiltext{3}{Centre for Astrophysics, Cosmology and Gravitation, Astronomy Dept, University of Cape Town, Rondebosch, South Africa, 7701}
\altaffiltext{4}{Department of Astronomy, New Mexico State University, Las Cruces, NM 88003}
\altaffiltext{5}{Department of Astronomy, University of Virginia, Charlottesville, VA 22903}
\altaffiltext{6}{The Observatories, Carnegie Institution of Washington, Pasadena, CA
91101}
\altaffiltext{7}{Dept. of Physics \& Astronomy, Michigan State University, East Lansing, MI 48824}
\altaffiltext{8}{Visiting Astronomer, Kitt Peak National Observatory, National
Optical Astronomy Observatory, which is operated by the Association of
Universities for Research in Astronomy, Inc., under cooperative agreement with
the National Science Foundation.}
\altaffiltext{9}{South African Astronomical Observatory, Observatory, South Africa, 7935}
\altaffiltext{10}{Institute of Astronomy, University of Vienna, A-1180, Vienna, Austria}
\altaffiltext{11}{STScI, Baltimore MD 21218}
\altaffiltext{12}{Harvard-Smithsonian Center for Astrophysics,  Cambridge MA 02138}
\altaffiltext{13}{Department of Physics and Astronomy, Open University, Milton Keynes UK MK7 6AA}
\altaffiltext{14}{Department of Physics and Astronomy, Uppsala University, 75120, Uppsala, Sweden}
\altaffiltext{15}{Department of Physics and Astronomy, Brigham Young University, Provo, UT 84602}
\altaffiltext{16}{Department of Physics and Astronomy, Dartmouth College, Hanover, NH 03755}
\altaffiltext{17}{Visiting Astronomer, Cerro-Tololo Inter-American Observatory, which is operated by the Association of
Universities for Research in Astronomy, Inc., under cooperative agreement with
the National Science Foundation.}
\altaffiltext{18}{Institute of Astrophysics, Georg-August-University, Friedrich-Hund-Platz 1, D-37077, G\"ottingen, Germany}
\altaffiltext{19}{Department of Radiotherapy, Medical University of Vienna, A-1090 Vienna, Austria}



\begin{abstract}
We present new absolute trigonometric parallaxes and proper motions for seven Pop II variable stars: five RR Lyr variables;  RZ Cep, XZ Cyg, SU Dra,  RR Lyr, UV Oct; and two type 2 Cepheids; VY Pyx and \kp. We obtained these results with astrometric data from Fine Guidance Sensors, white-light interferometers on {\it Hubble Space Telescope}. We find absolute parallaxes  in milliseconds of arc: RZ Cep, \RZpi mas; XZ Cyg, \XZpi mas; SU Dra, \SUpi mas;  RR Lyr, \RRpi mas; UV Oct, \UVpi mas; VY Pyx, \VYpi mas; and \kp ,  \kppi mas; an average $\sigma_\pi/\pi$ = 5.4\%. 
With these parallaxes we compute absolute magnitudes in V and K bandpasses corrected for interstellar extinction and Lutz-Kelker-Hanson bias. Using these RRL absolute magnitudes, we then derive zero-points for 
M$_V$-[Fe/H] and M$_K$-[Fe/H]-Log\,P relations. The technique of reduced
parallaxes corroborates these results. We employ our new results
to determine distances and ages of  several Galactic globular clusters
and the distance of the LMC.  The latter is close to that previously
derived from Classical Cepheids uncorrected for any metallicity effect,
indicating that any such effect is small.
We also discuss the somewhat puzzling results obtained for our two type 2
Cepheids.

\end{abstract}




%

 \section{Introduction}

RR Lyrae variable stars (RRL) have long played a crucial role in
understanding old stellar populations (Pop II). Paraphrasing Smith (1995)\nocite{Smi95},
they are important as tracers of the chemical and dynamical properties
of old populations, as standard candles in our own and nearby galaxies,
and as a test bed for the understanding of stellar pulsation and
evolution. Their luminosities are of great potential importance
in estimating the distances and hence the ages of globular clusters
--both the absolute ages and the relative ages as a function of
metallicity, [Fe/H]. An error in distance modulus of 0.1 magnitude corresponds
to an age uncertainty of 1 Gyr. The RRL are also vital for studies of the structure and formation of our Galaxy, Local Group members and other nearby galaxies, a field which is currently referred to as Near-Field Cosmology. Their importance as distance indicators comes from the fact that
they follow M(V)-[Fe/H] and K-log\,P 
or K-[Fe/H]-log\,P
relations. 
The zero points of these relations have been much discussed.
Trigonometric parallaxes remain the only fundamental method of getting
RRL distances and luminosities, free of the assumptions which go
into other methods discussed in Section~\ref{MV} below.
Absolute parallaxes allow these assumptions to be tested.
What is required is an improved fundamental zero-point
calibration. which currently rests on the \HST~parallax \citep{Ben02a} of RR Lyrae alone (c.f. Sollima \etal 2006).
In this paper we apply the astrometric precision of {\it HST}/FGS to the determination of absolute parallaxes for five galactic RRL: XZ Cyg = Hip 96112; UV Oct = Hip 80990; RZ Cep = Hip 111839; SU Dra = Hip 56734; RR Lyr = Hip 95497; and two type 2 Cepheids, \kp = Hip 93015 and VY Pyx = Hip 43736. 
Target properties are given in Table~\ref{tbl-AP} and discussed in Section~\ref{OBS}. 

Type 2 Cepheids (hereafter CP2), more luminous than the RR Lyraes, have great potential as distance
indicators in old populations. They have recently been shown to
define a narrow K-band Period-Luminosity Relation (Matsunaga \etal  2006, 2009) \nocite{Mat06, Mat09}
with little metallicity dependence.  The slope and zero-point of this relation are
indistinguishable from that of the RRL derived by Sollima \etal (2006).\nocite{Sol06}
Two CP2, $\kappa$ Pavonis and VY Pyxidis (confirmed as such by \nocite{Zak00} Zakrzewski \etal 2000), were expected to be
sufficiently close that  very accurate parallaxes and absolute magnitudes
could be obtained with \HST. Not only could these parallaxes, likely a factor of three more precise than from {\it HIPPARCOS}, provide an accurate zero-point
for the CP2 Period-Luminosity Relation (PLR), but they may facilitate the derivation of the slope and zero-point
of a combined RRL and CP2 PLR. \cite{Maj10} has recently asserted that CP2 and RRL define a single-slope PLR when a Wesenheit magnitude, W$_{VI}$=V-2.45(V-I)  is plotted against log\,P. 


In the following sections we describe our astrometry using one of our targets, \kp, as an example throughout. This longest-period member of our sample has been identified as a peculiar W Vir star \citep{Fea08} but, if included,  could anchor our K-band PLR slope.  Hence, its parallax value deserves as much external scrutiny as possible. We discuss (Section~\ref{OBS}) data acquisition and analysis; present the results of spectrophotometry of the astrometric reference stars required to correct our relative parallax to absolute (Section~\ref{SpecPhot}); derive absolute parallaxes for these  variable stars (Section~\ref{PAR}); derive  absolute magnitudes (Section~\ref{MAGS}); determine (Section~\ref{LL}) a K-band PLR zero point and an M$_V$-[Fe/H] relation zero point, and compare our resulting absolute magnitudes with past determinations; in Section~\ref{DSA} discuss the distance scale ramifications of our results, and apply our PLR zero points to two interesting classes of object -  Globular Clusters (M3, M4, M15, M68, $\omega$ Cen, and M92) and the LMC. In Section~\ref{CP2sec} we discuss the puzzling results for VY Pyx and \kp. We summarize  our findings in Section~\ref{SUMM}.

\section{Observations and Data Reduction}  \label{OBS}

\cite{Nel10}
provides an overview of the
Fine Guidance Sensor (FGS) instrument (a two-axis shearing interferometer), and \cite{Ben07} describe the fringe tracking (POS) mode astrometric capabilities 
of an FGS, along with the data acquisition and reduction strategies also used in the present study. 
We time-tag our data with a modified Julian Date, MJD  =  JD - 2400000.5.

Between thirteen and twenty-three sets of astrometric data were acquired with {\it HST} FGS 1r for each of our seven science targets.  We obtained most of these sets at epochs determined by field availability, primarily dictated by two-gyro guiding constraints. See \cite{Ben10} for a brief discussion of these constraints. 
The various complete data aggregates span from 2.37 to 13.14 years. Table~\ref{tbl-LOO} contains the epochs of observation, pulsational phase, the V magnitude, and estimated B-V color index (required for the lateral color correction discussed in Section \ref{ASTMOD}) for each variable. 
The B-V colors are inferred from phased color curves 
constructed from various sources: XZ Cyg \citep{Stu66}; RZ Cep \citep{Epp73}; SU Dra \citep{Bar02}; RR Lyr \citep{Har55}; UV Oct (Kolenberg, private comm.); \kp~\citep{Sho92}; VY Pyx \citep{San91}.

\begin{figure}
\epsscale{1.0}
\plotone{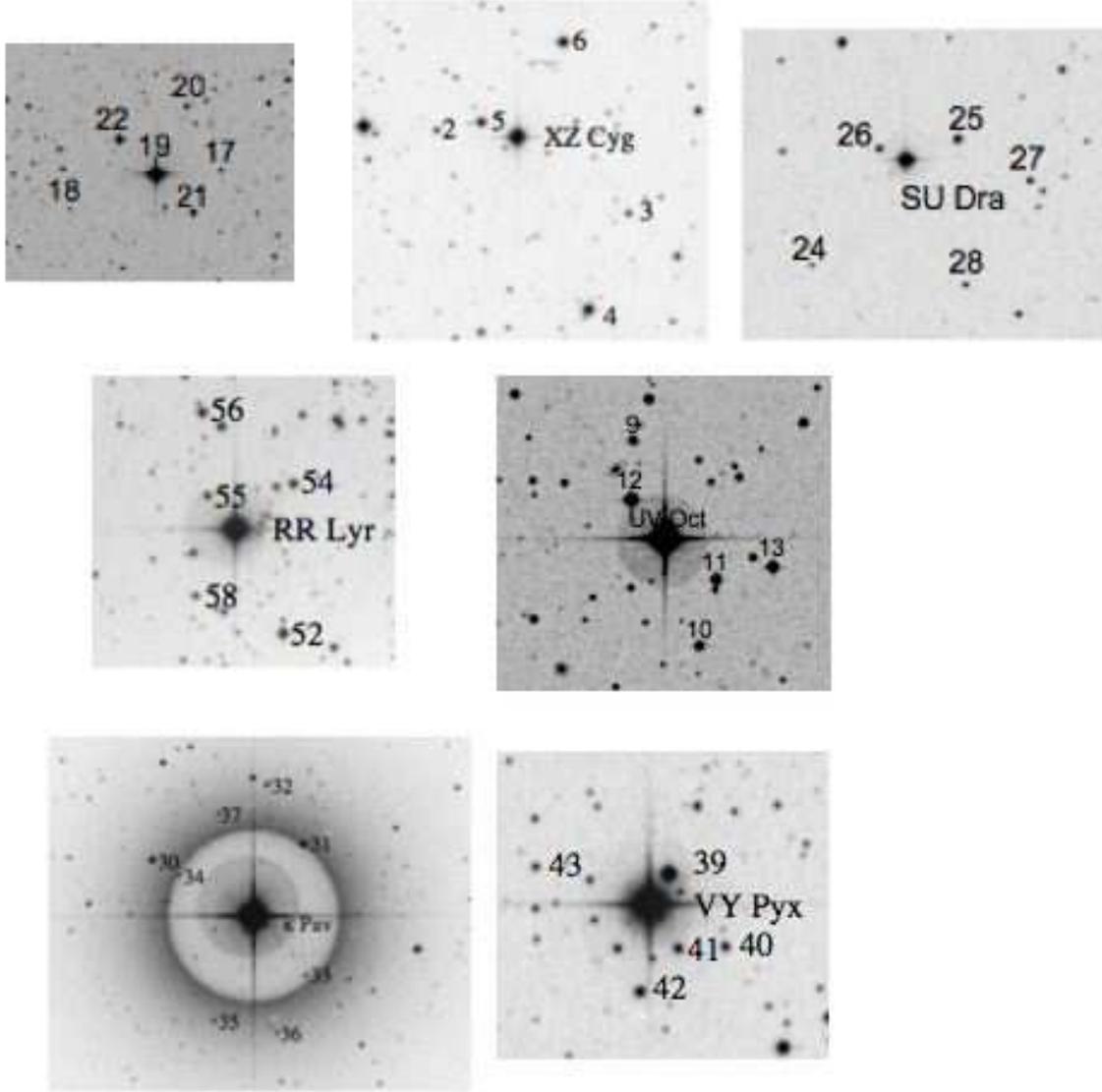}
\caption{The RRL and CP2 fields with astrometric reference stars marked. Boxes are roughly 2\arcmin ~across with North to the top, East to the left. RZ Cep is at top left.}
\label{Fig1}
\end{figure}

Each individual \HST~data set required approximately 33 minutes of spacecraft time. The data were reduced and calibrated as detailed in  McArthur \etal (2001)\nocite{mca01}, Benedict \etal (2002a)\nocite{Ben02a}, Benedict \etal (2002b)\nocite{Ben02b}, Soderblom \etal (2005), and \cite{Ben07}. At each epoch we measured reference stars and the target multiple times to correct for intra-orbit drift of the type seen in the cross filter calibration data shown in figure 1 of Benedict \etal (2002a)\nocite{Ben02a}.  The distribution of reference stars  on a second generation Digital Sky Survey R image near each of our science targets is shown in Figure \ref{Fig1}. 
The orientation of each successive observation changes, mandated by {\it HST} solar panel illumination constraints. 

Data are downloaded from the \HST~archive and passed through a pipeline processing system. This pipeline extracts the astrometry measurements (typically one to two minutes of fringe x and y position information acquired at a 40 Hz rate, which yields several thousand discrete measurements), extracts the median (which we have found to be the optimum estimator of position), corrects for the Optical Field Angle Distortion \nocite{McA02}(McArthur \etal 2002), and attaches all required time tags and parallax factors.  

Table~\ref{tbl-AP} collects measured properties for our target variables, including stellar type (ab or c for RRL), log of the pulsational period, $\langle$V$\rangle$, $\langle$K$\rangle$, $\langle$B-V$\rangle$, E(B-V), A$_V$, and A$_K$. Photometry is from the various sources noted in the table. The $\langle$K$\rangle$ is in the 2MASS system.  All reddening values are  adopted from those listed in \cite{Fer98b} or \cite{Fea08} with a sanity check provided by our reference star photometry. 

Our default metallicity source is  \cite{Fer98b}. The metallicity of RR Lyr is from \cite{Kol10}. The metallicities of UV Oct and VY Pyx were determined for this paper, using the approach described in \cite{Kol10}, determined by analysis of Fe line equivalent widths measured from
high-resolution spectra. The \kp~metallicity is from \cite{Luc89}.  The \cite{Fer98b} metallicities agree with \cite{Lay94} for the brighter stars in common (V $<$ 11). Because the \cite{Lay94} metallicities are on the ZW  \citep{Zin84} scale we assume the same scale for the \cite{Fer98b} metallicities. Because the \cite{Kol10} RR Lyrae metallicity agrees with \cite{Fer98b}, we presume that it too is ZW. Therefore we believe our metallicities are on, or close to, the ZW scale. 
This is the scale we use to establish zero-points that will be applied later to derive distances. 

Finally, three stars in our sample (including RR Lyrae) exhibit Blazhko cycles, wherein
the maximum and minimum brightness vary over time. Smith and Kolenberg have
studied this phenomenon for many such stars (see, e.g., LaCluyz«e et al. 2004\nocite{LaC04}, Kolenberg
et al. 2006, 2010\nocite{Kol06, Kol10}, Blazhko Project website http://www.univie.ac.at/tops/blazhko/) and conclude
from recent data that the total output of the target Blazhko stars averaged over a cycle
remains constant within 0.03 mag, in accordance with the findings by \cite{Alc03}.
As the peak brightness decreases, the minimum brightness increases. Blazhko is not a disease
that renders RRL poor standard candles as further demonstrated in \cite{Cac05}.

\setcounter{footnote}{0}
\section{Spectrophotometric Parallaxes of the Astrometric Reference Stars} \label{SpecPhot}
The following review of our astrometric and spectrophotometric techniques uses the \kp~field as an example. 
Because the parallaxes determined for the variables will be
measured with respect to reference frame stars which have their own
parallaxes, we must either apply a statistically derived correction from relative to absolute parallax (Van Altena, Lee \& Hoffleit 1995, hereafter YPC95) \nocite{WvA95} or estimate the absolute parallaxes of the reference frame stars.  In principle, the colors, spectral type, and luminosity class of a star can be used to estimate the absolute magnitude, M$_V$, and V-band absorption, A$_V$. The absolute parallax is then simply,
\beq
\pi_{abs} = 10^{-{(V-M_V+5-A_V)}\over5}
\eeq

The luminosity class is generally more difficult to estimate than the spectral type (temperature class). However, the derived absolute magnitudes are critically dependent on the luminosity class. As a consequence we use as much additional information as possible in an attempt to confirm the luminosity classes. Specifically, we obtain 2MASS\footnote{The Two Micron All Sky Survey
is a joint project of the University of Massachusetts and the Infrared Processing
and Analysis Center/California Institute of Technology } photometry and  proper motions from the PPMXL catalog \citep{Roe10} for a one degree square field centered on each science target, and iteratively employ the technique of reduced proper motion (Yong \& Lambert 2003\nocite{Yon03}, Gould \& Morgan 2003\nocite{Gou03}) to confirm our giant/dwarf classifications (Section~\ref{MODCON}). 

\subsection{ Reference Star Photometry}
Our band passes for reference star photometry include: BV from recent measurements with the New Mexico State University 1m telescope \citep{Hol10} for RR Lyr, SU Dra, XZ Cyg, and  RZ Cep fields; from the South African Astronomical Observatory (SAAO) 1m for the UV Oct, 
\kp~and VY Pyx fields; from the SMARTS 0.9m \citep{Sub10} for  the VY Pyx and UV Oct fields; and JHK (from 2MASS).  
Table \ref{tbl-IR} lists BVJHK photometry for our reference stars bright enough to have 2MASS measurements. 

\subsection{Reference Star Spectroscopy }
Spectral classifications for reference stars in the UV Oct, 
\kp~and VY Pyx fields were provided by the South African Astronomical Observatory (SAAO) 1.9m telescope.  The SAAO resolution  
was 3.5 \AA/ (FWHM) with wavelength coverage from 3750 \AA$\leq$ $\lambda$ 
$\leq$ 5500 \AA.  Spectroscopic classification of the reference stars in the fields of
RR Lyr, SU Dra, XZ Cyg, and  RZ Cep was accomplished using data obtained with the
Double Imaging Spectrograph (DIS) on the Apache Point Observatory 3.5 m 
telescope\footnote{The
Apache Point Observatory 3.5 m telescope is owned and operated by
the Astrophysical Research Consortium.}. We used the high-resolution gratings, delivering a dispersion 
of 0.62 \AA /pix, and covering the wavelength range of 3864 $\leq$ $\lambda$ 
$\leq$ 5158 \AA. Spectroscopy of the reference stars in the fields of UV Oct, 
\kp~and VY Pyx was also obtained using the RC Spectrograph on
the CTIO Blanco 4 m. The Loral3K CCD detector with KPGL1-1 grating was used to
deliver a dispersion of 1.0 \AA /pix, covering the wavelength range
3500 $\leq$ $\lambda$ $\leq$ 5830 \AA. 
Classifications used a combination of template matching and line ratios. Spectral types for the stars are generally better than $\pm$2
subclasses. 

\subsection{Interstellar Extinction} \label{AV}
\begin{figure}
\epsscale{1.00}
\plotone{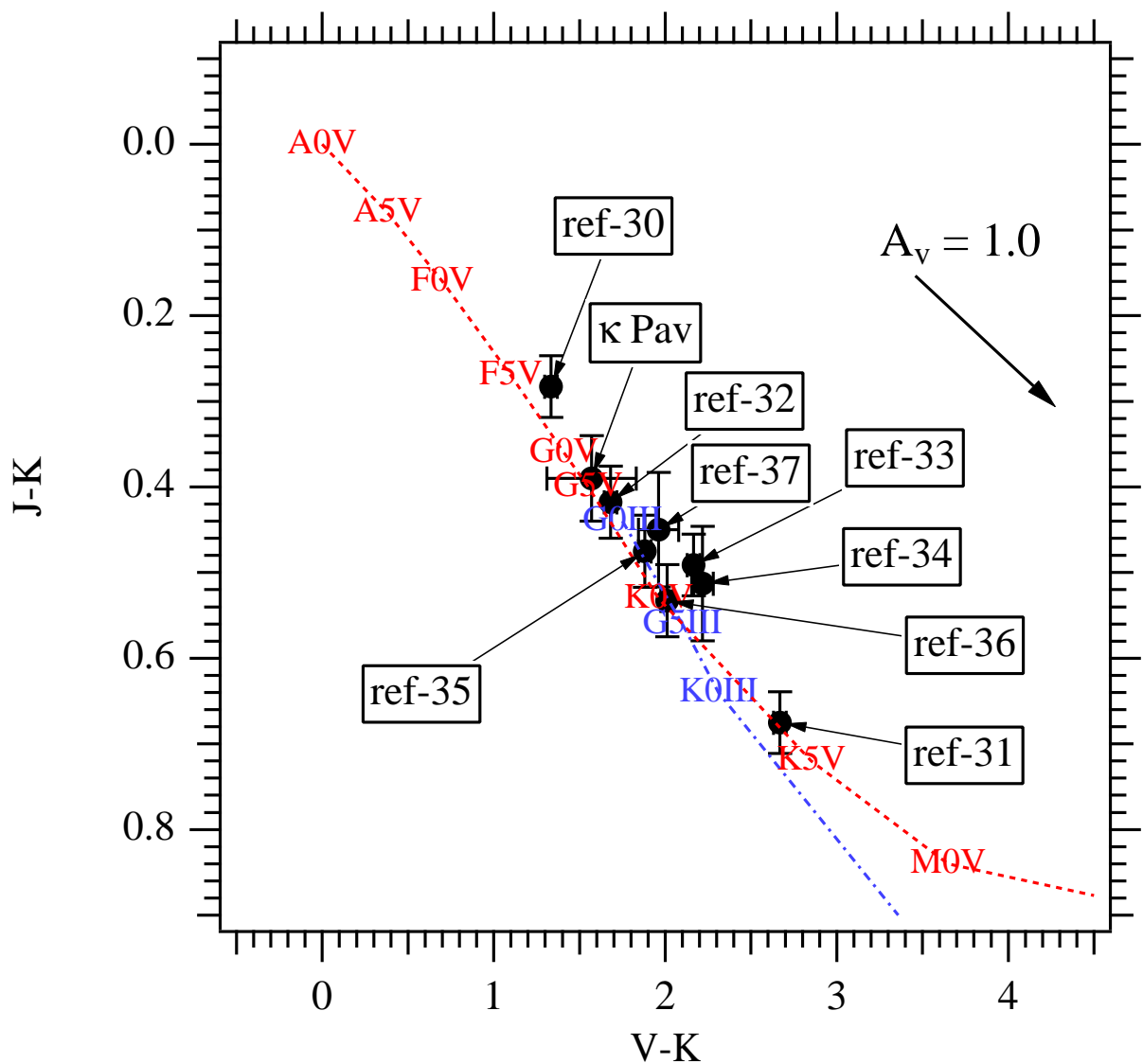}
\caption{J-K vs V-K color-color diagram for \kp~and reference stars. The dashed line is the locus of  dwarf (luminosity class V) stars of various spectral types; the dot-dashed line is for giants (luminosity class III). The reddening vector indicates A$_V$=1.0 for the plotted color systems. For this field at Galactic latitude $\ell^{II}=-25\arcdeg$, $\langle A_V\rangle$ = 0.05 $\pm$ 0.06 magnitude (Table~\ref{tbl-SPP}) with a maximum of 0.22 \citep{Sch98}.}
\label{Fig2}
\end{figure}

To determine interstellar extinction we first plot the reference stars on a J-K vs. V-K color-color diagram. A comparison of the relationships between spectral type and intrinsic color against those we measured provides an estimate of reddening. Figure \ref{Fig2} contains the \kp~J-K vs V-K color-color diagram and reddening vector for A$_V$ = 1.0. Also plotted are mappings between spectral type and luminosity class V and III from Bessell \& Brett~(1988)\nocite{Bes88} and Cox (2000)\nocite{Cox00}. Figure~\ref{Fig2}, along with the estimated spectral types, provides an indication of the reddening for each reference star. 

Assuming an R = 3.1 Galactic reddening law (Savage \& Mathis 1979\nocite{Sav79}), we derive A$_V$ values by comparing the measured colors (Table \ref{tbl-IR} ) with intrinsic (V-K)$_0$ and (B-V)$_0$  colors from  Cox (2000). We estimate A$_V$ from A$_V$ = 1.1E(V-K) = 3.1E(B-V), where the ratios of total to selective extinction were derived from the Savage \& Mathis (1979) reddening law and a reddening estimate in the direction of \kp~from \cite{Sch98}, via NED\footnote{NASA/IPAC Extragalactic Database}.  All resulting A$_V$ are collected in Table~\ref{tbl-SPP}. These are the A$_V$ used in Equation 1. 
 
Using the \kp~field as an example, we find that the technique of reduced proper motions can provide a possible confirmation of reference star estimated luminosity classes. The precision of existing proper motions for all the reference stars is $\sim$5 mas y$^{-1}$, only suggesting discrimination between giants and dwarfs. Typical errors on H$_K$, a parameter equivalent to absolute magnitude, M, were about a magnitude. Nonetheless, a reduced proper motion diagram did suggest that ref-31 is not a dwarf star. 
Our luminosity class uncertainty is reflected in the input spectrophotometric parallax errors (Table~\ref{tbl-SPP}). We will revisit this additional test in Section~\ref{MODCON}, once we have higher precision proper motions obtained from our modeling.

\subsection{Estimated Reference Frame Absolute Parallaxes}

We derive absolute parallaxes for each reference star using M$_V$ values as a function of spectral type and luminosity class from \cite{Cox00} and the A$_V$ derived from the photometry. Our adopted errors for (m-M)$_0$ are 0.5 mag for all reference stars. This error includes uncertainties in A$_V$ and the spectral types used to estimate M$_V$.
Our reference star parallax estimations from Equation 1 are listed in Table \ref{tbl-SPP}. Similar data for the RR Lyr reference frame can be found in \cite{Ben02a}.
For the \kp~field individually, no reference star absolute parallax is better determined than ${\sigma_{\pi}\over \pi}$ = 23\%. The average absolute parallax for the reference frame is $\langle\pi_{abs}\rangle = 1.5$ mas.
We compare this to the correction to absolute parallax discussed and presented
in YPC95. Entering
YPC95,  section 3.2, fig. 2, with the \kp~Galactic
latitude, $\ell$ = -25\arcdeg, and average magnitude for the
reference frame, $\langle$V$_{ref}$$\rangle$= 14.2, we obtain a correction
to absolute of 1.2 mas. This gives us confidence in our spectrophotometric determination of the correction to absolute parallax. As in past investigations we prefer to introduce into our reduction model our spectrophotometrically estimated reference star parallaxes as observations with error. The
use of spectrophotometric parallaxes offers a more direct (less Galaxy model-dependent) way of
determining the reference star absolute parallaxes.

\section{Absolute Parallaxes of Population II Variable Stars}\label{PAR}

\subsection{The Astrometric Model} \label{ASTMOD}

With the positions measured by FGS 1r (and FGS 3 for RR Lyr) we determine the scale, rotation, and offset ``plate
constants" relative to an arbitrarily adopted constraint epoch (the so-called ``master plate") for
each observation set (the data acquired at each epoch). The MJD of each observation set is listed in Table~\ref{tbl-LOO}, along with a measured magnitude transformed from the FGS instrumental system as per Benedict \etal (1998)\nocite{Ben98a}, but with coefficients determined for FGS 1r. Our \kp~reference frame contains 6 stars. Several primary science targets (RR Lyr, VY Pyx, and \kp) are bright enough to require the use of the FGS neutral density filter.  For those objects we use the modeling approach outlined in Benedict \etal (2002b)\nocite{Ben02b}, with corrections for both cross-filter and lateral color positional shifts, using values specific to FGS 1r or FGS 3 determined from previous calibration observations with each FGS. 

We employ GaussFit (Jefferys \etal 1988)\nocite{Jef88} to minimize $\chi^2$. The solved equations
of condition for the \kp~ field are:
\beq
        x'  =  x + lc_x(\it B-V) {- \Delta XFx}
\eeq
\beq
        y'  =  y + lc_y(\it B-V) { - \Delta XFy}
\eeq
\beq
\xi = Ax' + By' + C  - \mu_x \Delta t  - P_\alpha\pi_x
\eeq
\beq
\eta = -Bx' + Ay' + F  - \mu_y \Delta t  - P_\delta\pi_y
\eeq

where $\it x$ and $\it y$ are the measured coordinates from {\it HST};
$\it lc_x$ and $\it lc_y$ are the
lateral color corrections; $\Delta$XFx and $\Delta$XFy are the cross filter corrections in $\it x$ and $\it y$, applied only to the observations of RR Lyr and the CP2; and $\it B-V $ are
the  B-V  colors of each star.  A  and B   
are scale and rotation plate constants, C and F are
offsets; 
$\mu_x$ and $\mu_y$ are proper motions; $\Delta$t is the epoch difference from the mean epoch;
$P_\alpha$ and $P_\delta$ are parallax factors;  and $\it \pi_x$ and $\it \pi_y$
 are  the parallaxes in x and y. We obtain the parallax factors from a JPL Earth orbit predictor (Standish 1990\nocite{Sta90}), upgraded to version DE405. 

\subsection{Prior Knowledge and Modeling Constraints} \label{MODCON}
\begin{figure}
\epsscale{0.75}
\plotone{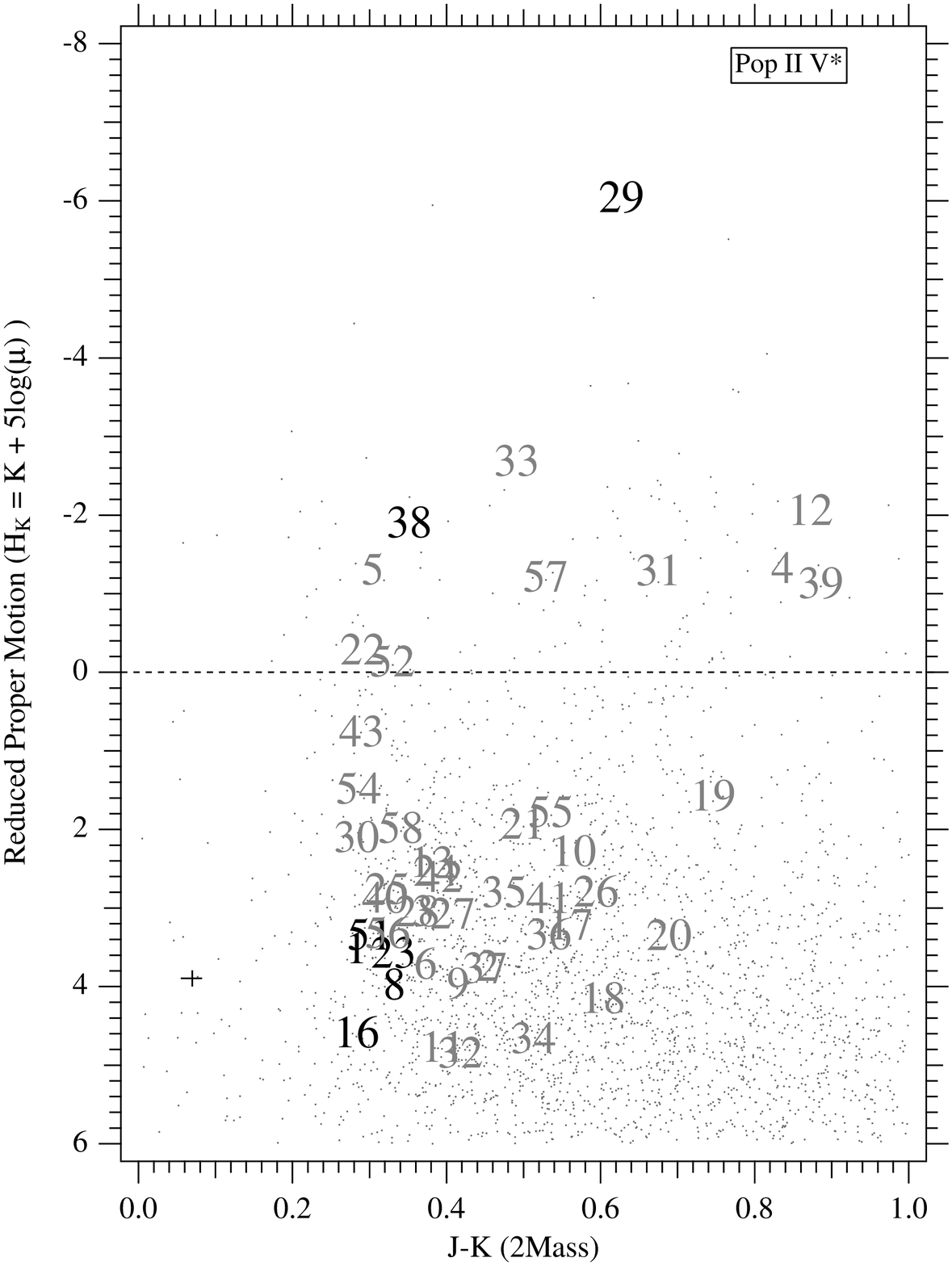}
\caption{Reduced proper motion diagram for 4039 stars taken from ${1\over3}$\arcdeg $\times$ ${1\over3}$\arcdeg~fields centered on each variable star. Star identifications are shown (black) for XZ Cyg (1), UV Oct (8), RZ Cep (16), SU Dra (23), \kp~(29), VY Pyx (38), RR Lyr (51), and for (grey) all astrometric reference stars in Table~\ref{tbl-SPP}. Ref-52 through -58 are from \cite{Ben02a}. H$_K$ for all numbered stars is calculated using our final proper motions, examples of which for the \kp~field can be found in Table~\ref{tbl-PM}. For a given spectral type giants and sub-giants have more negative H$_K$ values and are redder than dwarfs in J-K. Reference stars ref-4, -12, -31, -39 are confirmed giants. The plotted position (but not the colors from Table~\ref{tbl-IR}) suggests a sub-giant classification for ref-33. The cross in the lower left corner indicates representative internal errors along each axis.} \label{Fig3}
\end{figure}

In a quasi-Bayesian approach the reference star spectrophotometric absolute parallaxes (Table~\ref{tbl-SPP}) and PPMXL proper motions (Table~\ref{tbl-PM}) 
were input as observations with associated errors, not as hardwired quantities known to infinite precision.  Input proper motion values  have typical errors of 4--6 mas y$^{-1}$ for each coordinate.  The lateral color and cross-filter calibrations and the B-V color indices are also treated as observations with error. Proper motion values obtained from our modeling of {\it HST} data for the \kp~field are listed in Table~\ref{tbl-PM}. Transverse velocities for \kp~and all our other science targets, given our final parallaxes, are listed below. We employ the technique of reduced proper motions to provide a confirmation of all reference star estimated luminosity classes listed in Table~\ref{tbl-SPP}. We obtain proper motion and J, K photometry from PPMXL and 2MASS for a ${1\over3}$\arcdeg $\times$ ${1\over3}$\arcdeg~field centered on all RRL and CP2. Figure~\ref{Fig3} shows H$_K$ = K + 5log($\mu$) plotted against J-K color index for 4039 stars. If all stars had the same transverse velocities, Figure~\ref{Fig3} would be equivalent to an HR diagram. The RRL, CP2, and associated reference stars are plotted as ID numbers from Table~\ref{tbl-PM}. With our now measured, more precise proper motions (Table~\ref{tbl-PM}) errors in H$_K$ are now $\sim0.3$ magnitude. Note the clumping of the RRL towards the 'faint' end of the diagram. Reduced proper motion diagrams are 'fooled' by the relatively high space velocities of these halo component giant stars.

We stress that for no CP2 or RRL in our program was a previously measured parallax used as prior knowledge and entered as an observation with error. Only reference star parallax prior knowledge was so employed. Our parallax results are blind to previous RRL and CP2 parallax measures from \HIP~ and/or parallaxes from surface brightness estimates.

\subsection{Assessing Reference Frame Residuals}
\begin{figure}
\epsscale{0.5}
\plotone{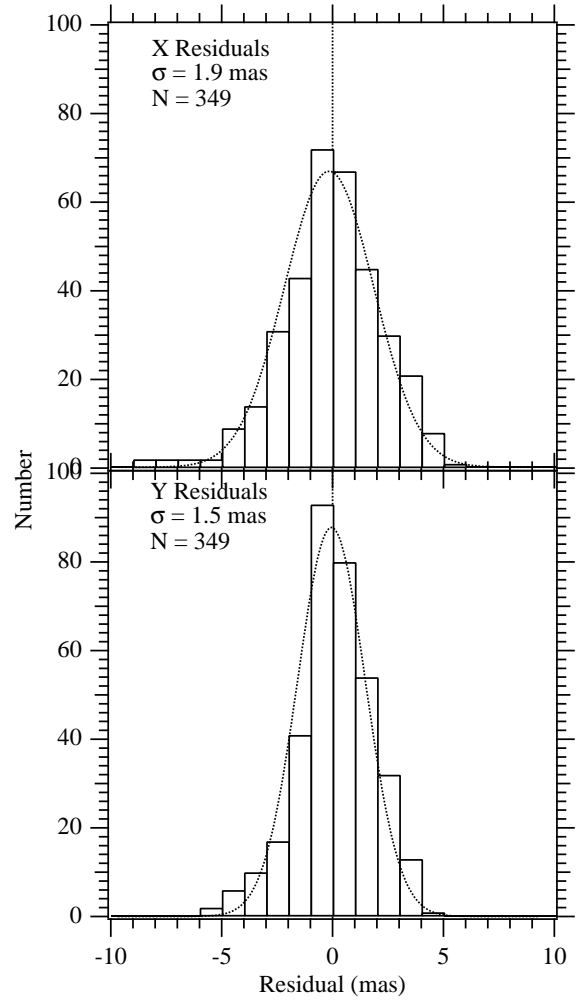}
\caption{Histograms of x and y residuals obtained from modeling \kp~and astrometric reference stars with equations 4 and 5, constraining D=-B and E=A. Distributions are fit with gaussians whose 1-$\sigma$ dispersions are noted in the plots.} \label{Fig4}
\end{figure}

The Optical Field Angle Distortion calibration (McArthur \etal 2002\nocite{McA02}) reduces as-built {\it HST} telescope and FGS 1r distortions with amplitude $\sim1\arcsec$ to below 2 mas over much of the FGS 1r field of regard.  From histograms of the \kp~field astrometric residuals (Figure~\ref{Fig4}) we conclude that we have obtained satisfactory correction. The resulting reference frame `catalog' in $\xi$ and $\eta$ standard coordinates (Table \ref{tbl-POS}) was determined
with	average position errors $\langle\sigma_\xi\rangle= 0.46$	 and	$\langle\sigma_\eta\rangle = 0.46$ mas.

To determine if there might be unmodeled - but possibly correctable -  systematic effects at the 1 mas level, we plotted  reference frame X and Y residuals against a number of spacecraft, instrumental, and astronomical parameters. These included X, Y position within our total field of view; radial distance from the field of view center; reference star V magnitude and B-V color; and epoch of observation.  We saw no obvious trends. 

\subsubsection{The Absolute Parallax of \kp} \label{AbsPi}
We constrain $\pi_x = \pi_y$ in Equations 3, 4 and obtain for \kp~a final absolute parallax $\pi_{abs} = 5.57 \pm0.28$ mas.  We have achieved a significant reduction in formal error, with the HIP97 determination, $\pi_{abs} = 6.00 \pm0.67$ mas and  the HIP07 determination, $\pi_{abs} = 6.52 \pm0.77$ mas.   A surface brightness (pulsation) parallax for \kp~was determined by \cite{Fea08} to be 4.90 $\pm$ 0.17 mas. The parallax of \kp~derived in the present work is in better agreement
with the pulsation parallax of \cite{Fea08} (a two sigma difference) than the
HIP97 and HIP07 parallaxes.
 We note that this object is another for which the HIP07 re-reduction has not improved agreement with \HST.  See \cite{Bar09} for a few other examples involving galactic Cepheids.
Parallaxes and relative proper motion results for all RRL and CP2  are collected in Tables~\ref{tbl-SUM1} and~\ref{tbl-SUM2}. 

\subsubsection{Modeling Notes on the RRL and VY Pyx}\label{ASTnotes}
Final model selection for all fields was based on reference star placement relative to the target, total number of reference stars, reduced $\chi^2$ ($\chi^2$/DOF, where DOF = degrees of freedom), and parallax error. For all but the the \kp , RR Lyr, and RZ Cep fields  we increased  the number of modeling coefficients in Equations 3 and 4 to six. We introduced radial terms, resulting in these equations of condition
\beq
\xi = Ax' + By' + G(x'^2+y'^2)^{1/2} + C  - \mu_x \Delta t  - P_\alpha\pi_x
\eeq
\beq
\eta = -Bx' + Ay' + H(x'^2+y'^2)^{1/2} + F  - \mu_y \Delta t  - P_\delta\pi_y
\eeq
  Absolute parallaxes, relative proper motions, and transverse velocities for \kp~and associated reference stars are collected in Table~\ref{tbl-PM}. Parallaxes for all RRL and CP2 are collected in Tables~\ref{tbl-SUM1} and \ref{tbl-SUM2}.

All our absolute parallaxes directly rely on the estimates of reference star parallaxes. Should anyone wish to  verify our results independently, the reference stars used in this study are all identified in archival material\footnote{\url http://www.stsci.edu/observing/phase2-public/11211.pro} held at the Space Telescope Science Institute. Adopted reference star spectral types for all fields are listed in Table~\ref{tbl-SPP}. 

{\bf XZ Cyg} - Reference star 2 was removed from the data set because of high and unmodelable residuals. Application of Equations 6,7 to the remaining data resulted in a positional catalog with $\langle\sigma_\xi\rangle= 0.28$ and $\langle\sigma_\eta\rangle = 0.29$ mas. Residuals histograms are well-modeled with Gaussians of dispersion $\sigma_x = 1.4$ mas and $\sigma_y = 1.1$ mas. 
The resulting parallax, $\pi_{abs} = 1.67 \pm0.17$ mas, agrees within the far larger errors of  either the HIP97 or HIP07 value, $\pi_{abs} = 2.29 \pm0.85$ mas.

{\bf UV Oct} - Application of Equations 6,7 to these data resulted in a positional catalog with $\langle\sigma_\xi\rangle= 0.19$ and $\langle\sigma_\eta\rangle = 0.18$ mas. Residuals histograms are well-modeled with Gaussians of dispersion $\sigma_x = 1.0$ mas and $\sigma_y = 0.8$ mas.  The resulting parallax, $\pi_{abs} = 1.71 \pm0.10$ mas is between the HIP97 ($\pi_{abs} = 1.48 \pm0.94$ mas) or HIP07 ($\pi_{abs} = 2.44 \pm0.81$ mas) values.

{\bf RZ Cep} - Reference star 20 was removed from the data set because of high residuals. Application of Equations 4,5 to these data resulted in a positional catalog with $\langle\sigma_\xi\rangle= 0.37$ and $\langle\sigma_\eta\rangle = 0.37$ mas. Residuals histograms are well-modeled with Gaussians of dispersion $\sigma_x = 1.7$ mas and $\sigma_y = 1.2$ mas. 
The resulting parallax, $\pi_{abs} = 2.54 \pm0.19$ mas differs from the HIP97 ($\pi_{abs} = 0.22 \pm1.09$ mas) or HIP07 ($\pi_{abs} = 0.59 \pm1.48$ mas) values. The \HST~value has a far smaller error.

{\bf SU Dra} - Reference star 27 was removed from consideration because of high residuals. Application of Equations 6,7 to these data resulted in a positional catalog with $\langle\sigma_\xi\rangle= 0.34$ and $\langle\sigma_\eta\rangle = 0.39$ mas. Residuals histograms have Gaussians with dispersion $\sigma_x = 0.9$ mas and $\sigma_y = 1.0$ mas.  The resulting parallax, $\pi_{abs} = 1.42 \pm0.16$ mas agrees within the larger errors  of both HIP97 ($\pi_{abs} = 1.11 \pm1.09$ mas) and HIP07 ($\pi_{abs} = 0.20 \pm1.13$ mas), but is far more statistically significant.

{\bf RR Lyr} - Because temporary onboard science-side failures left \HST~with few operational science instruments in late 2008, we were granted additional orbits for FGS astrometry. One of our targets was RR Lyr, a field for which we obtained five additional orbits. Application of Equations 4,5 to our original FGS 3 and these new data resulted in a positional catalog with $\langle\sigma_\xi\rangle= 0.34$ and $\langle\sigma_\eta\rangle = 0.52$ mas. Residuals histograms are well-fit with Gaussians of dispersion $\sigma_x = 0.7$ mas and $\sigma_y = 0.7$ mas. 
The resulting parallax, $\pi_{abs} = 3.77 \pm0.13$ mas is between the HIP97 ($\pi_{abs} = 4.38 \pm0.59$ mas) or HIP07 ($\pi_{abs} = 3.46 \pm0.64$ mas) values. Our previous parallax value  \citep{Ben02a} was  $\pi_{abs} = 3.82 \pm0.20$ mas. The additional \HST~data have significantly improved the parallax and proper motion precision compared to the 2002 values.

\begin{figure}
\epsscale{1.0}
\plotone{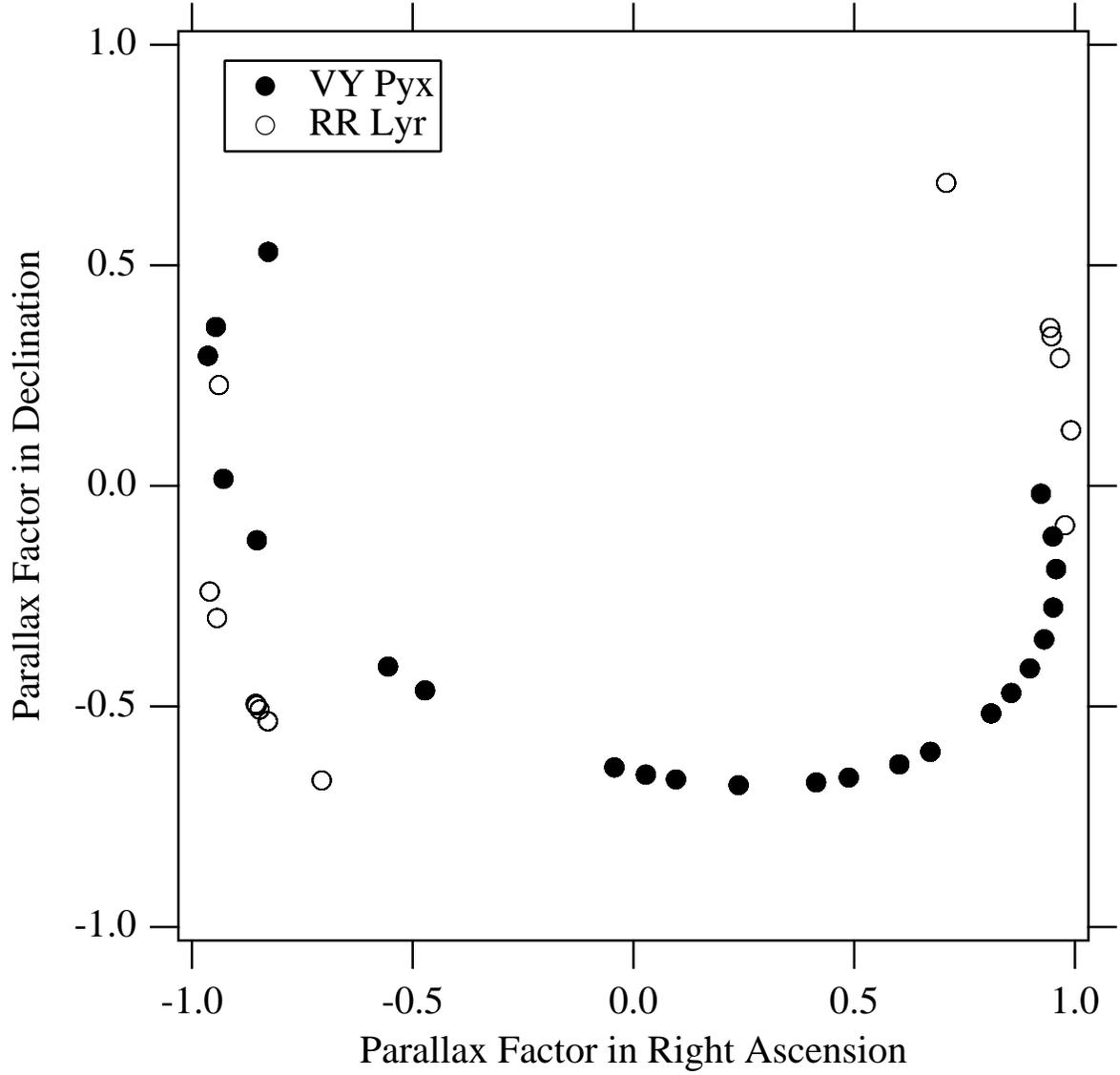}
\caption{Sampling of the parallactic ellipses of VY Pyx and RR Lyr. Lack of coverage is not an issue in the parallax of VY Pyx.} \label{Fig5}
\end{figure}

{\bf VY Pyx} - Application of Equations 4,5 to these data resulted in a positional catalog with $\langle\sigma_\xi\rangle= 0.23$ and $\langle\sigma_\eta\rangle = 0.22$ mas. Residuals histograms are well-modeled with Gaussians of dispersion $\sigma_x = 1.4$ mas and $\sigma_y = 1.0$ mas.  The resulting parallax, $\pi_{abs} = 6.44 \pm0.23$ mas is larger than either the HIP97 ($\pi_{abs} = 5.74 \pm0.76$ mas) or HIP07 ($\pi_{abs} = 5.01 \pm0.44$ mas) values. We assessed the residuals from our modeling for evidence of orbital motion that could impact a parallax determination and found no significant signals. As done for all our modeling, we tested each spectrophotmetrically determined reference star parallax by solving for a trigonometric parallax relative to the aggregate of reference stars and found no significant departures from the initial estimates.  One last potential impact on a final parallax would be inadequate sampling of the parallactic ellipse. We show in Figure~\ref{Fig5} the parallax factor coverage for both VY Pyx and RR Lyr. We are confident that our parallax is not affected by poor sampling of the parallactic ellipse of VY Pyx. We shall discuss this parallax result later in Section~\ref{MAGS} when we determine absolute magnitudes and in Section~\ref{CP2sec}, wherein we discuss both these peculiar CP2.

\subsection{{\it HST} Parallax Accuracy}
Our parallax precision, an indication of our internal, random error, is $\sim$ 0.2 mas. To assess our accuracy, or external error, we  have compared (Benedict \etal 2002b,  Soderblom \etal 2005) our parallaxes with results from independent measurements from {\it HIPPARCOS} (Perryman \etal 1997).  See \cite{McA11} for a more recent comparison with the {\it HIPPARCOS} re-reduction of \cite{Lee07a}. Other than for the Pleiades (Soderblom \etal 2005), we  have no large systematic differences with {\it HIPPARCOS} for any objects with ${\sigma_{\pi}\over\pi}<$10\%. The next significant improvement in geometrical parallaxes for Pop II variable stars will come from the space-based, all-sky astrometry missions {\it Gaia} \citep{Lin08}  with $\sim20~ \mu$arcsec precision parallaxes. Final results are expected early in the next decade.

\section{The Absolute Magnitudes of the RRL and CP2} \label{MAGS}

In using measured quantities involving distance, care has to be taken of bias questions.
\cite{Lut73} in a well known paper used a frequentist argument to show that
if stars of the same measured parallax are grouped together, the derived mean parallax
will be overestimated.  This is because
for most Galactic stellar distributions, the stars with overestimated parallaxes will out number those with
underestimated parallaxes.  This argument can be applied to single stars chosen by parallax and the
argument can be put in a Bayesian form (see for example section 5 of
Benedict et al. 2007). \nocite{Ben07}
There have been extensive discussions of the method in the literature (see e.g Smith 2003)\nocite{Smi03} . Here we have
used the general formulation of Hanson (1979) \nocite{Han79} as applied to the determination of absolute magnitudes. This Lutz-Kelker-Hanson (LKH) bias in absolute magnitude is proportional to $(\sigma_{\pi}/\pi)^2$. Presuming that all RRL and CP2 in Table 2 belong to the same class of object  (evolved Pop II stars), we scale the LKH correction determined in \cite{Ben02a} for RR Lyr and obtain the LKH bias corrections listed in Tables~\ref{tbl-SUM1} and~\ref{tbl-SUM2}. The average LKH bias correction for all objects in this study is -0.047 magnitude. We identify the choice of prior for this bias correction as a possible contributor to systematic errors in the zero-points of our PLR at the 0.01 magnitude level. For our example target, \kp, we find LKH = -0.02 magnitude (Table~\ref{tbl-SUM2}). We have used these corrected absolute magnitudes in deriving zero-points of the relations
discussed below. In addition we have used the uncorrected parallaxes to derived these zero-points by the method of reduced parallaxes (RP). This RP approach avoids some of the bias problems (see Feast 2002 whose general
scheme we use).\nocite{Fea02}

\subsection{An Absolute Magnitude for \kp} \label{kppec}
With $\langle$V$\rangle$= 2.78  (Table~\ref{tbl-AP}) and  given the  absolute parallax, \kppi mas from Section \ref{AbsPi}, we determine a distance modulus for \kp. 
For all objects (except RZ Cep, where we adopt the Fernley et al. 1998 value) we adopt a color excess from \cite{Fea08}, which for \kp~(and an adopted R = ${\rm{A}_V \over E(B-V)} = 3.1$) yields $\langle$A$_V$$\rangle$ = 0.05. With this $\langle$A$_V\rangle$, the measured distance to \kp, and the LKH correction we obtain M$_V = -1.99\pm 0.11$ and a corrected true distance modulus,  (m-M)$_0$ = 6.29.  From the value in \cite{Fea08}, $\langle$K$_s\rangle$ = 2.78 we obtain M$_K= -3.52\pm 0.11$. 

\kp~has been identified to be a peculiar W Vir star \citep{Fea08, Fea10}. See Section~\ref{CP2sec} for additional discussion. 
Results, including all proper motions and absorption- and LKH bias-corrected absolute magnitudes, for the objects in our program  are collected for the  CP2 in Table~\ref{tbl-SUM2} and for the RRL in Table~\ref{tbl-SUM1}.

\section{Zero-points for the RRL Period-Luminosity and M$_V$ - Metallicity Relations}\label{LL}

\subsection{The RRL M$_K$ - log\,P Relation}\label{KPLR}
A relation between K and log P for RRL was found by \cite{Lon86}
in globular clusters. In more recent times a number of such relations have been suggested and these are given in Equations~\ref{K1}-\ref{K6}, where the zero-points a$_n$ refer to the mean period and metallicity of our parallax sample.
 The log\,P of RZ Cep, an overtone type `c' RRL, has been `fundamentalized' by adding +0.127, a factor determined by comparing type ab and type c RRL, e.g. \cite{Oas06}. Equation~\ref{K1} was obtained by Sollima (2008) from globular clusters with distance based on subdwarf parallaxes.
Since the metallicity term is small we also give the equation without the metallicity term (Equation~\ref{K2}). Equation~\ref{K3} is a semi-theoretical derivation. Equation~\ref{K4} is from RRL in the cluster Reticulum in the LMC.
Equation~\ref{K5} was derived from RRLs of different metallicities in the globular cluster  $\omega$ Cen,
and Equation~\ref{K6} is from RRL in the field of the LMC. For these relations the metallicities are all
on, or close to the Zinn-West system, except  Equation~\ref{K1} where they are on the Carretta-Gratton system. 
The relations between different systems provided by \cite{Car97} show that our mean 
ZW metallicity (-1.58) converts to  -1.40 on the CG scale, and in view of the small size of
the metallicity coefficient in Equation~\ref{K1} the effect is negligible ($\sim 0.01$ mag).

The values of a$_n$ for these equations are listed in Table~\ref{tbl-KZP}. Two values are given for each equation, one derived
by fitting the LKH corrected absolute magnitudes to the equation, and one derived using the method of
reduced parallaxes (RP). The difference in the two zero-points are within the uncertainties.
Figure~\ref{Fig6} shows a K- log\,P plot for our data. A slope of  -2.38 (Equation~\ref{K2}) was adopted for the fitted line.
The CP2 \kp , included in the plot, will be discussed below (Section~\ref{CP2sec}). It is not included in the fit.
The various PLR relationships and their sources are:
\begin{equation}
M_{K} = (-2.38\pm0.04)(\log P +0.28) + (0.08\pm0.11) ([Fe/H] + 1.58) + a_{1},~{\rm Sollima~et~al.~2008}  \label{K1}
\end{equation}
\begin{equation}
M_{K} = (-2.38\pm0.04) (\log P +0.28) + a_{2},~{\rm Sollima~et ~al. ~2006, neglecting~metallicity} \label{K2}
\end{equation}
\begin{equation}
M_{K} = -2.101(\log P +0.28) + (0.231\pm0.012)([Fe/H] +1.58) + a_{3},~{\rm Bono~et~al.~2003}   \label{K3}
\end{equation}
\begin {equation}
M_{K} = (-2.16\pm0.09)(\log P +0.28) + a_{4},~{\rm Dall'Ora~et~al.~2004}  \label{K4}
\end{equation}
\begin{equation}
M_{K} = (-2.71\pm0.12)(\log P +0.28) + (0.12\pm0.04)([Fe/H] + 1.58) + a_{5},~{\rm Del~Principe~et~al.~2006}  \label{K5}
\end{equation}
\begin{equation}
M_{K} = (-2.11\pm0.17)(\log P +0.28) + (0.05\pm0.07)([Fe/H] + 1.58) + a_{6},~{\rm Borissova~et~al.~2009}  \label{K6}
\end{equation}

\subsection{An RRL M$_V$ - [Fe/H] Relation Zero-Point from \HST~Parallaxes}
There is a long history of attempts to determine how M$_V$
depends on [Fe/H]. A linear relation is generally assumed.
Our data hint at a slope with the more metal-poor stars brighter.
The best estimate for the slope (b) is probably from the work of
\cite{Gra04}, using RRL in the LMC  (b = 0.214), and we have
adopt that slope here. 
Figure~\ref{Fig7} presents  our M$_V$ plotted against metallicity, [Fe/H], where the metallicity measures are from the sources noted in Table~\ref{tbl-AP}, all on the ZW scale. 
Fitting the function  \citep{Gra04}
\begin{equation}
M_V = (0.214\pm0.047)([Fe/H] +1.5) +a_7
\end{equation}
to all RRL, we obtain a zero-point, $a_7 $= +0.45 $\pm$ 0.05, listed in Table~\ref{tbl-KZP}. Hence, M$_V =+0.45 \pm 0.05$ for RRL with [Fe/H]=-1.50. The regression was carried out using GaussFit \citep{Jef88}, which takes into account uncertainties along both axes. The RMS of this fit, 0.08 mag in M$_V$, suggests an upper limit on the V-band cosmic dispersion in the absolute magnitudes of RRL. An RP approach finds M$_V =+0.46 \pm 0.03$.  Note that the  mean metallicity of our five RRL,  $\langle$ [Fe/H] $\rangle$=-1.58, is so close to -1.50,
that the error in the slope makes no significant difference to the zero-point. \cite{Bon07} find for field RRL a quadratic expression relating M$_V$ to [Fe/H]; M$_V\propto$ 0.50[Fe/H]+ 0.09[Fe/H]$^2$. We fit (again with GaussFit) the distribution seen in Figure~\ref{Fig7}, constraining  the [Fe/H] coefficients to the Bono \etal values and find a zero-point a = 0.98$\pm$0.05. This and our average $\langle$[Fe/H]$\rangle$=-1.58 yields $\langle$M$_V$$\rangle$=0.42, consistent with the \cite{Gra04} parameterization.

Regarding the intrinsic dispersion in RRL absolute magnitudes due to evolutionary effects, the intrinsic width of the RRL distribution in GC near this metallicity has been shown by \cite{San90} to be $\pm$0.2 mag. Thus the standard deviation of a uniformly filled strip is 0.057. For five stars, the standard error we would find given such a strip (absent observational uncertainty) is 0.029 mag. Our observational uncertainty standard error for the five is 0.024 mag. Combine the two in quadrature and our claim of +/-0.05 mag is actually a bit conservative.

\subsection{Comparison with Previous Determinations of RRL M$_V$ and M$_K$}\label{MV}
We compare in Table~\ref{tbl-MvComp} past determinations of RRL M$_V$  with our new value, M$_V =+0.45 \pm 0.05$ from LKH and M$_V =+0.46 \pm 0.03$ from the RP determination. An historical
summary to the early 1990s is given by Smith (1995). Carretta \etal (2000b)\nocite{Car00b}, Cacciari \& Clementini (2003)\nocite{Cac03}, and Di Criscienzo \etal (2006) \nocite{DiC06} discuss more
recent results.

Four methods have generally been applied to the problem of Pop II
distances: trigonometric parallaxes, cluster-based distances,  statistical parallaxes, and surface
brightness analyses.    In  the following we
will use the absolute magnitude at [Fe/H] = -1.5 to inter-compare methods. 
\begin{itemize}
\item {\bf Trigonometric Parallaxes} are essentially free from complicating
assumptions. However, one must still consider whether
the instability region is sufficiently populated by the few data
to give an unbiased relation. The first reliable parallax of an
RRL came from the {\em HIPPARCOS} satellite. The 
\cite{Koe98} result is from an analysis of Hipparcos parallaxes of RRL.
\cite{Fer98b} used the
parallax for RR Lyr itself (with an adopted [Fe/H] = -1.39) of $4.38\pm0.59$ mas to estimate 
M$_V=0.78\pm0.29$ mag.  {\em HIPPARCOS} also determined a distance to our target CP2 with  10--13\% errors. There are now two versions of the \HIPs catalog: \cite{Per97} (HIP97) and \cite{Lee07a} (HIP07). 
We have reanalyzed the field for RR Lyr itself,  which benefitted from additional FGS data secured in late 2008.  The \cite{Ben02a} \HST~parallax provided M$_V$=0.61$\pm$0.11, compared to our new value, M$_V$=0.54$\pm$0.07. Because the new and old parallaxes agree within their respective errors (Section~\ref{ASTnotes}), we ascribe the difference in M$_V$ primarily to a newer and presumably more accurate extinction determination, A$_V$ =0.13 \citep{Kol10}.  HIP97 and HIP07 parallaxes can be compared with the present \HST~results in Tables~\ref{tbl-SUM2} and \ref{tbl-SUM1}.

\cite{Gra98} derived an RRL scale based on the Hipparcos parallaxes of field horizontal branch stars.
The value in  Table~\ref{tbl-MvComp}   is a slight update of this result \citep{Car00b}.

\item {\bf Cluster-based distances}  generally rely upon  main sequence or
horizontal branch fitting calibrated by stars with well-determined distances.
 Using \HIP~ parallaxes
of 56 subdwarfs and data for 9 globular clusters, Carretta et al. (2000) \nocite{Car00b} obtain distances and hence 
the absolute magnitudes of the RRLs the clusters contain. Their result (Table~\ref{tbl-MvComp}) agrees well with ours,
though their errors are large. Thus our results promise improved accuracy in cluster distances.

\item{\bf Statistical Parallaxes} are derived by combining proper motions and radial velocities.   Fernley et al. (1998b) \nocite{Fer98b} performed such an analysis based on the Hipparcos proper motions.  In an elaborate reanalysis of those data, Gould \& Popowski (1998)  \nocite{Gou98} confirmed the Fernley et al. result with a slightly smaller uncertainty.  We quote the Gould \& Popowski value in Table 10.
There is about a 2 sigma difference from our value. This difference is not in itself of very high
significance. However the consequences of adopting RRLL absolute magnitudes 0.3 mag brighter than that
suggested by the statistical parallaxes is highly important for distance scale applications.
The reason for the fainter result obtained from statistical parallaxes is not clear. However
the statistical work depends on the adoption of a Galactic model. The result  may be due to
deviations from the adopted model (due possibly to stellar streams in the Galactic Halo).
A full analysis of the Galactic motions of RRLs based on our new absolute magnitude scale is
desirable. 

It is, however, interesting to note the work of  Martin \& Morrison (1998) \nocite{Mar98} who calculate the mean space motion (using radial velocities
and proper motions)
of Halo RRL with respect to the Sun as a function of the adopted absolute magnitude
(their fig 6). The velocity (V) in the direction of galactic rotation is quite
sensitive to the absolute magnitude. For our derived value of the absolute magnitude
scale their results predicts V =-250 \kms. The  galactic rotational velocity
(with respect  to the Sun) was recently determined from VLBI of Galactic masers
\citep{Bru11} as 246$\pm$ 7 \kms. Our absolute magnitude scale implies
that the halo RRL form a non-rotating system. Earlier values of the galactic rotation were smaller, and our result would have implied
a retrograde halo.

Lastly, \cite{Dam09} obtained a calibration of a K-log\,P relation from statistical parallaxes which yields an M$_K$ $\sim 0.4$ mag
fainter than our result. This suggests that the difference in M$_V$, comparing statistical parallaxes and
our trigonometric work, is  not due to problems with corrections for interstellar absorption.

\item{\bf Surface Brightness} Extending the early efforts of Baade and Wesselink, Barnes \& Evans (1976) introduced a technique for determining pulsating star distances using differential surface brightness measurements. \nocite{Bar76}  There are two main uncertainties in the application of this general method to RRL.
First, shocks occurring in the atmospheres of the stars during part of the
pulsation cycle complicate the interpretation of the radial velocity curves. Secondly,
it is necessary to adopt a value of p (the ratio of pulsation velocity to measured
radial velocity) and this remains uncertain. Also, not all color indices are equally
effective in predicting a surface brightness. In addition to a result based on  many RRL from \cite{Fer98a}, the present status of the RRL
M$_V$ from this method is captured in the paper by \cite{Cac03}, who have introduced a number of refinements,
but their work is for only one star, RR Cet. 
Our parallaxes may produce a more accurate {\em p} factor for RRL, and may  refine {\em p} for CP2.

\end{itemize}


\section{Distance Scale Applications}\label{DSA}
We now apply our new zero-points to several globular clusters and the LMC. The globular clusters were chosen because they had existing RRL photometry in both the V- and the K-band. We also estimate globular cluster ages with these new distance moduli. All parameterizations, sources of slopes and zero-points, and distance moduli derived from our new RRL absolute magnitudes listed in Table~\ref{tbl-SUM1} and plotted in Figures~\ref{Fig6} and  \ref{Fig7} are summarized in Table~\ref{tbl-DGC} (Globular Clusters) and Table~\ref{tbl-DLMC} (LMC). In each case we assume that A$_V$=3.1E(B-V), and A$_K$=0.11A$_V$. 

\subsection{Distance Moduli of Globular Clusters}\label{GCDM}
\begin{figure}
\epsscale{0.85}
\plotone{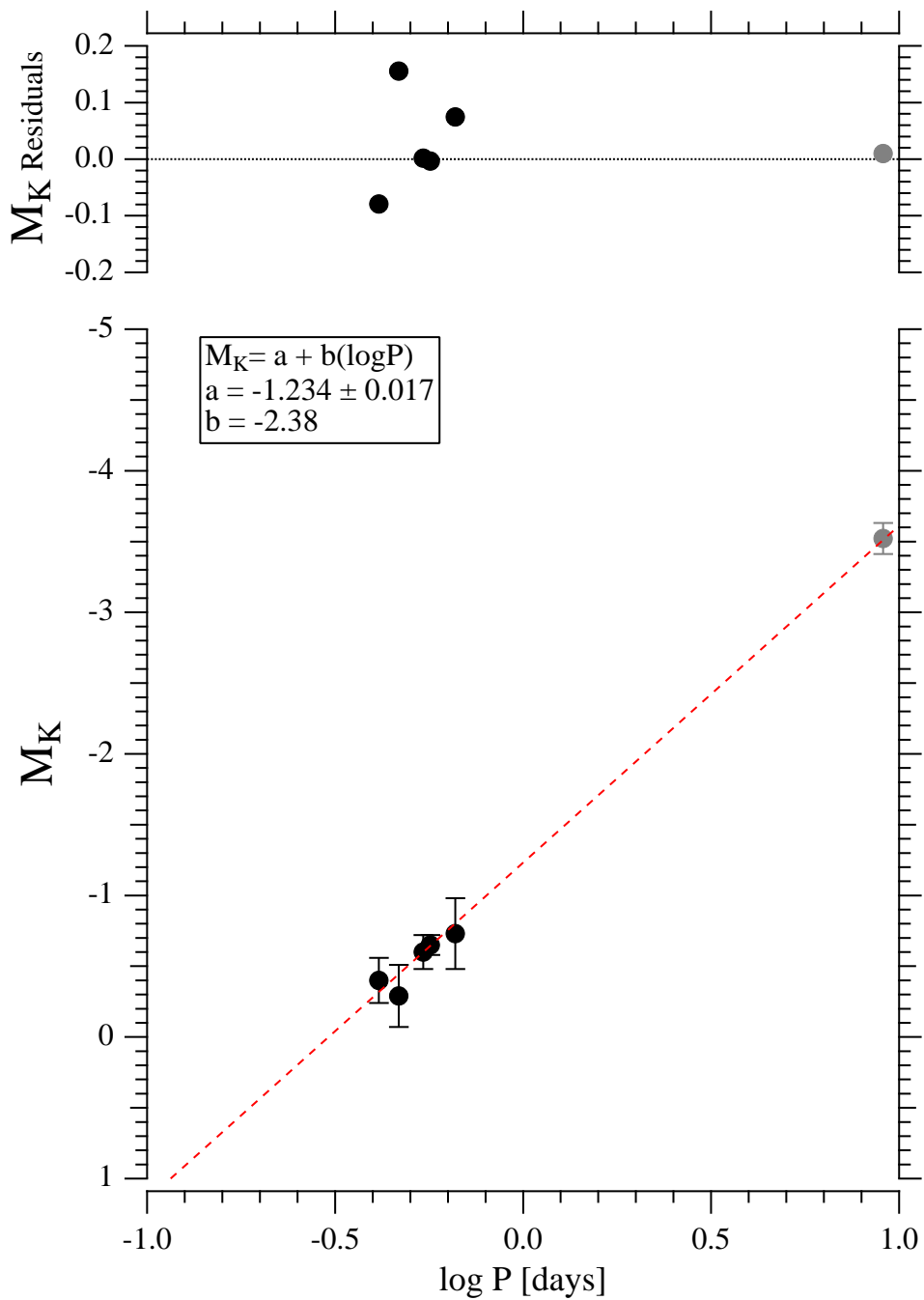}
\caption{A Pop II K-band PLR. All magnitudes have been corrected for interstellar extinction. Coefficients are for M${_K} = a +b*(logP)$. Zero-point (a) error is $1\sigma$. The absolute magnitude (uncorrected for `peculiarity') and residual for \kp~are plotted in grey.  The fit is without  \kp. The slope is constrained to b=-2.38 \citep{Sol06}. The largest residual is for XZ Cyg.} \label{Fig6}
\end{figure}

\begin{figure}
\epsscale{0.75}
\plotone{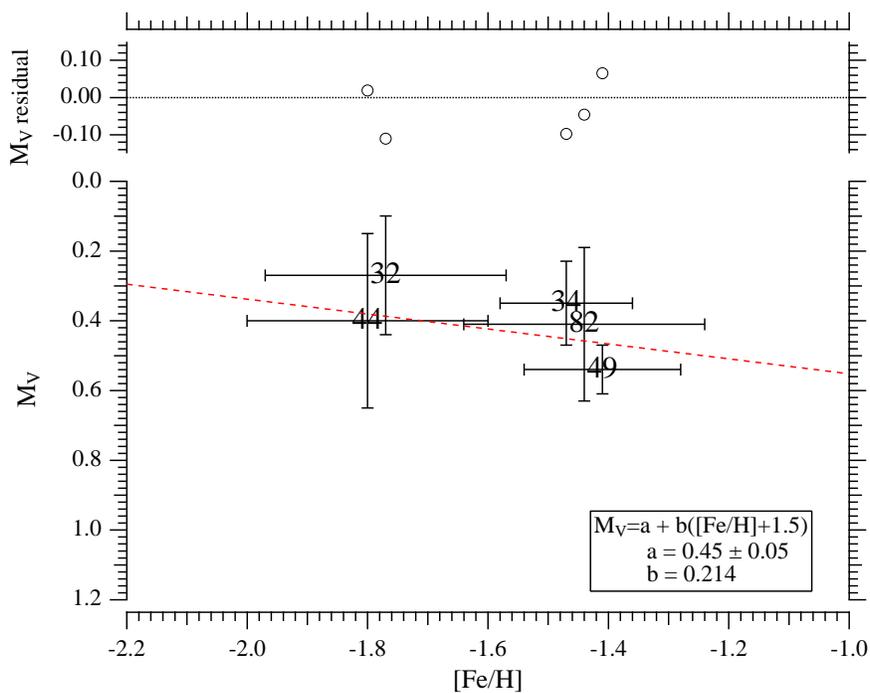}
\caption{RRL extinction-corrected absolute magnitude M$_{V}$ plotted against metallicity, [Fe/H], whose sources are given in Table~\ref{tbl-AP}. Objects  are identified by model number: RZ Cep =32; UV Oct = 34; SU Dra = 44; RR Lyr = 49; XZ Cyg = 82. Errors in M$_{V}$ and [Fe/H] are 1$-\sigma$. The dashed line is an impartial fit to both the absolute magnitude and metallicity data with an adopted slope, b = 0.214 \citep{Gra04}, resulting in a zero-point, a = +0.45 $\pm$ 0.05.  The RMS residual to this fit is 0.08 magnitudes.} \label{Fig7}
\end{figure}

A sample of six globular clusters was selected based on the availability of both K-band and V-band RRL photometry. We first employ the zero-point from M$_K$ versus log\,P, our Figure~\ref{Fig6},  comparing with the apparent magnitude PLR for each cluster. To each we also apply the M$_V$ - [Fe/H] relation shown in Figure~\ref{Fig7}, transforming the relevant [Fe/H] from \cite{Sol06} on the CG scale to the ZW scale, using the CG-ZW mapping established by \cite{Car97} (their equation 7). The final M$_V$ error includes the $\pm0.047$ magnitude M$_V$ - [Fe/H] slope error. The error in our adopted K-band PLR slope makes a negligible contribution to the final distance modulus error. The expectation is that the K and V distance moduli collected in Table~\ref{tbl-DGC} should agree. In all cases the two approaches yield the same distance modulus within the errors.

{\bf M3}  From \cite{Bnk06} we extract for RRL $\langle$V$_0 \rangle$=15.62$\pm$0.05, corrected for an assumed A$_V$=0.03. The Figure~\ref{Fig7} M$_V$ - [Fe/H] and an [Fe/H]=-1.34 (CG),  [Fe/H]=-1.57 (ZW) provide M$_V$=0.45$\pm$0.05, thus (m-M)$_0$= 15.17$\pm$0.12. The \cite{But03} K-band apparent magnitude PLR zero-point is 13.93$\pm$0.04. This, combined with our Figure~\ref{Fig6} zero-point yields (m-M)$_0$=15.16$\pm$0.06. 

{\bf M4}  For RRL \cite{Cac79}  find $\langle$V$_0 \rangle$=12.15$\pm$0.06, corrected for an assumed A$_V$=1.19. The Figure~\ref{Fig7} M$_V$ - [Fe/H] and an [Fe/H]=-1.40 (ZW) provide M$_V$=0.47$\pm$0.05, thus (m-M)$_0$= 11.68$\pm$0.13. The \cite{Lon90} K-band apparent magnitude at log\,P=-0.3 (figure 1f) is K(0)= 11.10$\pm$0.06, corrected for A$_K$=0.13. This, combined with our Figure~\ref{Fig6} zero-point yields (m-M)$_0$=11.48$\pm$0.08. The two approaches barely yield the same distance modulus within the errors, possibly due to the high and uncertain extinction correction due to known differential reddening. We note that increasing to E(B-V)=0.415 (from the adopted 0.36) equalizes the two distance moduli at (m-M)$_0=11.46$.

{\bf M15}  From  \cite{Sil95}, table 6, we derive $\langle$V$_0\rangle$=15.51$\pm$0.02, corrected for an assumed A$_V$=0.30. This value comes only from the RRL ab stars. The Figure~\ref{Fig7} M$_V$ - [Fe/H] and an [Fe/H]=-2.16 (ZW) provide M$_V$=0.31$\pm$0.05, thus (m-M)$_0$= 15.20$\pm$0.09. The \cite{Lon90} K-band apparent magnitude at log\,P=-0.3 (their figure 1c) is K$_0$= 14.67$\pm$0.10, corrected for A$_K$=0.03. This, combined with our Figure~\ref{Fig6} zero-point yields (m-M)$_0$=15.18$\pm$0.11.  

{\bf M68}  From  \cite{Wal94} we obtain $\langle$V$_0\rangle$=15.51$\pm$0.01, corrected for an assumed A$_V$=0.13. The Figure~\ref{Fig7} M$_V$ - [Fe/H] and an [Fe/H]=-2.08 (ZW) provide M$_V$=0.33$\pm$0.05, thus (m-M)$_0$= 15.18$\pm$0.08. The \cite{Dal06} K-band apparent magnitude at log\,P=-0.2 (their figure 3) is K$_0$= 14.35$\pm$0.04, corrected for A$_K$=0.01. This, combined with our Figure~\ref{Fig6} zero-point yields (m-M)$_0$=15.10$\pm$0.06.  

{\bf $\omega$ Cen} \cite{Del06} (figure 4) provides $\langle$K$_0\rangle$=13.05$\pm$0.06 for RRL ab at log\,P=-0.2. At that log\,P the Figure~\ref{Fig6} PLR yields M$_K(0)$=-0.75$\pm$0.05. Hence, (m-M)$_0 = 13.80\pm0.08$. Adopting [Fe/H]=-1.84 (ZW), we obtain M$_V$(0)=+0.38 from Figure~\ref{Fig7}. RRL V-band photometry from \cite{Ole03} and A$_V$=0.36 \citep{Sol06} provide $\langle$V$_0\rangle$=14.20$\pm$0.02, and (m-M)$_0 = 13.82\pm0.09$. 
 
{\bf M92}  From  \cite{Kop01} we derive $\langle$V$_0\rangle$=15.01$\pm$0.08, corrected for an assumed A$_V$=0.08.  The Figure~\ref{Fig7} M$_V$ - [Fe/H] and an [Fe/H]=-2.16 (ZW) provide M$_V$=0.31$\pm$0.05, thus (m-M)$_0$= 14.70$\pm$0.11. The \cite{Del05} K-band apparent magnitude at log\,P=-0.19 is K$_0$= 13.86$\pm$0.04, corrected for A$_K$=0.01. This, combined with our Figure~\ref{Fig6} zero-point yields (m-M)$_0$=14.64$\pm$0.06. The two approaches  yield the same distance modulus within the errors.

Within a year or so we will have an independent check on these globular cluster distance moduli. \cite{Cha11} are using the FGS on \HST~to obtain parallaxes of 9 metal-poor ([Fe/H] $< -1.5$) main sequence stars. The \HST~parallaxes are expected to have accuracies similar to those achieved for this RRL project, leading to absolute magnitude uncertainties of $\pm$0.05 mag for a given star. These stars will be used to test metal-poor stellar evolution models and to determine main sequence fitting distances to a large number of low metallicity globular clusters, including those above. See \cite{McA11} for an example of the construction of a main sequence using only a few highly precise absolute magnitudes.


\subsection{Globular Cluster Ages} \label{GCA}
Adopting the K-band distance moduli from Table~\ref{tbl-DGC},  we calculate absolute ages for our selected globular clusters  using a Monte Carlo simulation  similar to that described by \cite{Cha98}.  
We did not estimate an age for $\omega$ Cen, as the cluster is very complex with multiple stellar populations, and not conducive to a simple age determination.
For each remaining globular cluster, 3000 sets of isochrones were generated using the Dartmouth  stellar evolution program \citep{Cha01, Bjo06, Dot08}.  The input parameters for each set of isochrones were randomly selected from their distribution function as discussed by \cite{Bjo06}.   A total of 21 different parameters 
were varied in the stellar evolution calculations, including the nuclear reaction rates, opacities, 
surface boundary conditions, mixing length and composition.  The [Fe/H] values used 
in the stellar models were selected from a Guassian distribution with a standard deviation of
0.15 dex and a mean based upon high resolution spectroscopic abundance analysis of 
FeI lines \citep{Car09} and FeII lines \citep{Kra03}.  These independent [Fe/H] measurements agree quite well with each for each of the globular clusters, and the mean of the two measurements were used in the stellar model calculations\footnote{The high resolution spectroscopic [Fe/H] determinations 
for each cluster differ somewhat from the ZW scale listed in Table~\ref{tbl-DGC}.  Those [Fe/H] values were selected to be on the same [Fe/H] system as the target parallax stars (which, in general, do not have high dispersion spectroscopic [Fe/H] determinations).  For the purposes of stellar model calculations, the consistency between the field stars and globular clusters stars is not an issue, rather
one is interested in the absolute [Fe/H] scale, which is best determined from high resolution 
spectroscopic studies.}.

The age of each globular cluster was determined using the absolute magnitude of the point on the subgiant branch which is 0.05 mag redder than the turn-off \citep{Cha96}.  Photometry in V and I for each globular cluster except M15 was obtained from P.B.\ Stetson's photometric standard fields\footnote{http://www4.cadc-ccda.hia-iha.nrc-cnrc.gc.ca/community/STETSON/standards/} and used to determine the apparent magnitude of the subgiant branch.  The Stetson database for M15 does not reach the main sequence turn-off in I.  For this cluster, we used the HST ACS photometry from \cite{And08}.   The V band distance modulus for  each cluster was determined using the true distance moduli (derived from the K band) and reddening listed in Table~\ref{tbl-DGC}.  Errors in the distance moduli were assumed to be Gaussian with the uncertainty given in Table~\ref{tbl-DGC}, with the exception of M4.  M4 has a fairly high reddening, and there is evidence for differential reddening across the clusters.   Estimates for absorption in the V band range from $A_V = 1.16$ mag (using E(B-V) = 0.36 and the extinction calculator from  McCall 2004\nocite{McC04}) to $A_V = 1.33$ mag \citep{Ric97}.  We elected to use  $A_V = 1.22\pm 0.08$ mag, which implies an uncorrected $(m - M)_V = 12.70\pm 0.11$ mag for M4. Lastly, the age error for M15 was derived from the smaller error on the V-band distance modulus.

We determined the ages (collected in Table~\ref{tbl-DGC}) of the clusters to be: M3 $10.8\pm 1.0$ Gyr; M4 $(11.1^{ -1.4}_{+1.7})$ Gyr; M15 $12.1\pm 1.0$ Gyr; M68 $12.4\pm 1.0$ Gyr; and M92 $13.1\pm 1.1$ Gyr.  The larger error in the age of M4 is due to the larger uncertainty in the V band distance modulus to this cluster.  

Our absolute ages are in reasonable agreement with previous estimates.  For example, \cite{diC10} found an absolute age of 11.0+/- 1.5 Gyr for M92, which  agrees within the uncertainties with our age.  The differences between our age estimate and that of \cite{diC10} is due 
our use of an updated nuclear reaction rate, and a different distance modulus.  
 \cite{diC10} used the older NACRE  rate for the $^{14}\mathrm{N}(p,\gamma )^{15}\mathrm{O}$ nuclear reaction. We used the updated value for this critical nuclear reaction rate from \cite{Mar08}, which yields globular cluster ages approximately 1 Gyr older than the reaction rate used by di Cecco et al. \cite{diC10} adopted a distance modulus of 14.74 (no error reported), which is 0.1 mag larger than the distance modulus derived in this work. An increase in the distance modulus by 0.1 mag will decreased derived ages by approximately 1 Gyr.  The distance modulus  adopted by \cite{diC10} was the  one that gave the best fit between their theoretical isochrones and the observed color-magnitude diagrams.  It depends critically on their transformations for theoretical luminosities and temperatures to observed  magnitudes and colors.  

In general, it is difficult to find absolute age determinations for globular clusters in the literature.  Most works focus on relative age determinations, and the errors in the age estimates do not include 
uncertainties in the stellar evolution models and isochrones.  Although they focused on relative age determinations, \cite{Sal02} carefully determined the ages of a large sample of globular clusters, 
including all of the clusters whose ages are determined in this paper.  The difference between 
our ages and those derived by \cite{Sal02} are within the errors:
M92 $0.3\pm 1.4\,$Gyr; M68 $1.2\pm 1.3\,$Gyr; M3 $-0.5 \pm 1.2\,$Gyr; M15: $0.3\pm 1.4\,$Gyr and 
M4 $0.1\pm 1.7\,$Gyr.  

M3 is the only cluster we find to have a younger age than that derived by \cite{Sal02}.  This is likely due to the fact that the distance modulus we derive for this cluster is of order 0.1 mag larger than previous estimates, leading to our determination of a relatively young age for this cluster. Such a young age is supported by the fact that  M3 has a relatively red horizontal branch morphology for its metallicity.  A detailed differential study of M3 and M13 (clusters with similar metallicities) by \cite{Rey01} found that M3 was $1.7\pm 0.7\,$Gyr younger than M13.   The biggest age difference between our work and \cite{Sal02} is for the globular cluster
M68.  \cite{Sal02} used $[\mathrm{Fe/H}] = -2.00$ for this cluster while we adopt $[\mathrm{Fe/H}] = -2.33\pm 0.15\,$dex  based upon more recent spectroscopic studies. Using $[\mathrm{Fe/H}] = -2.00$ 
in our stellar evolution models reduces our age estimate for M68 by 1.1 Gyr to 11.3 Gyr and leads to good agreement with the \cite{Sal02} age estimate of this cluster of 11.2 Gyr.

\subsection{LMC Distance Moduli} \label{LMCcomp}
In this section we derive the distance modulus of the LMC using our derived RRL absolute
magnitudes.
We then compare it with that derived from classical Cepheids whose absolute magnitudes
were also based on \HST~trigonometrical parallaxes \citep{Ben07}.

From observations of RRL in the LMC Gratton et al. (2004) derived the relation
\begin{equation}
V_{0}= (0.214\pm0.047)([Fe/H] +1.5) + 19.064\pm0.017
\end{equation}
with metallicities on the ZW scale.
This with the results in line 7 of Table~\ref{tbl-KZP} yields a  true distance modulus of 18.61$\pm$ 0.05
(Table~\ref{tbl-DLMC}) from the LKH approach. The OGLE team \citep{Sos03} found a mean value of $V_{0} = 18.90 \pm 0.02$ for
7110 RRab and c stars in the LMC. This is from the OGLEII survey. The mean RRL magnitude
(uncorrected for reddening) is not changed in the larger OGLEIII survey \citep{Sos09}.
Using the Gratton relation and adopting  a mean metallicity of -1.53 for the LMC from
 \cite{Bor09} together with the zero points of Table~\ref{tbl-KZP} leads to an LMC modulus of 
18.46$ \pm $0.06. The differences between these two results for the LMC modulus is primarily
due to the fact that  the two groups adopt different reddenings for the LMC objects
(see Clementini et al. 2003). \nocite{Cle03}

For the infrared we have used the \cite{Bor09} work which has individual 
values of [Fe/H] and incorporates earlier work. Using Equation~\ref{K6} with a zero point
corresponding to a$_6$= -0.54 they obtain an LMC modulus of 18.53. The LK 
corrected zero-point in Table~\ref{tbl-KZP} then shows that our modulus is 0.02 mag brighter.
We adopt $ 18.55\pm 0.05$ based on the uncertainty of our zero-point and the uncertainty
in the infrared data. The result using the reduced parallax zero-point is 18.53. The reddenings adopted by Borissova et al. were
means of values derived in a variety of ways. The uncertainties in these
values have a very small effect (of order 0.01 mag) on the derived distance
modulus.

In view of the sensitivity of the derived LMC modulus to the reddening when using
the relation in V, it seems best to give most weight to the determination
using the relation in K.
While we believe that this is the current
best mean distance to the LMC from RR Lyraes, it should be noted that the LMC
is sufficiently close that its depth structure is important. Thus strictly,
the result applies to the selection of stars studied by Borissova et al.
and the model of the LMC that they adopt. A similar remark applies to other
determinations.

\cite{Dal04} established a K-Band PLR in the Reticulum cluster associated
with the LMC. 
\begin{equation}
K_{0} = -2.16logP + 17.33 \pm 0.03
\end{equation}
This, together with  a$_{4}$ in Table~\ref{tbl-KZP}, leads  to a corrected distance modulus for the
cluster of $18.50 \pm 0.03$. As discussed by Dall'Ora et al. the relative
distance of this cluster and the main body of the LMC has not been well established.  Our result
suggests that any difference in distance is small.

In view of the above discussion we adopt 18.55$ \pm $0.05 (LKH method) or 18.53 (reduced parallax method)
as the best RRL distance to the LMC. This may be compared with the LMC modulus  
obtained by \cite{Ben07}  from classical Cepheids based on \HST~parallaxes of Galactic stars of this type.
Their results were (m-M)$_{0} = 18.50 \pm 0.04$ from a PL relation in W$_{VI}$; $18.52 \pm 0.06$ from a PL relation
in V$_{0}$: and $18.48 \pm 0.04$ from a PL relation in K$_{0}$. These results have not been corrected for the metallicity
difference between the LMC Cepheids and the Galactic calibrating stars. There has been much discussion as to whether or not a metallicity correction to Cepheid absolute magnitudes is necessary. The above results show that
any correction is small; at least between Cepheids of Galactic and LMC metallicities. This result agrees with the
theoretical discussion of Bono et al. (2010). \nocite{Bon10}

We note that recent work \citep{Lan09} on the absolute magnitude calibration of red clump
stars, which formerly led to a rather low modulus for the LMC, now gives 
(m-M)$_{0} = 18.47\pm 0.03$ in good agreement with the RRL and (uncorrected for metallicity) Cepheid
results.

\section{The Type 2 Cepheids}\label{CP2sec}
In this section we discuss the two CP2 for which we have
obtained parallaxes and absolute magnitudes. 
\cite{Mat06} established an M$_K$- log\,P relation for CP2s in globular clusters
which shows little evidence of a metallicity dependence. It was therefore originally
anticipated that the parallaxes of these two stars, \kp~ (log\,P = 0.958) 
and VY Pyx (log\,P = 0.093) could be used to establish a zero-point for this relation,
which appears to be continuous with a PL relation for RRLs.
Subsequent work \citep{Sos09, Mat09} 
has shown that in the LMC field
the situation is more complex than in globular clusters. Some CP2
with periods near that of \kp~lie above the PL relation,
and have distinctive light curves (peculiar W Vir stars). Many of these stars
are known to be binaries. It has now been suggested \citep{Fea08} that \kp~belongs to this class,
though it is not known to be binary and the classification remains uncertain.

The identification of a star as belonging to an older stellar population can be based on kinematics and/or metallicity.  The RRL transverse velocities, V$_t$ in Table~\ref{tbl-SUM1}, all suggest  an identification with the halo, or that these stars have no connection with the local stellar thin disk population. 
As summarized in \cite{Maa07}, the identification of CP2 is more complex. First, there is separation by period and metallicity. The prototype short-period CP2 is BL Her with P=1.31$^d$ and [Fe/H]=-0.1 \citep{Maa07}. The long-period prototype CP2 is W Vir with P=17.27$^d$ and [Fe/H]=-1.0 \citep{Bar71}.  In addition to detailed metallicity variations by species the classification of CP2 also rests on distance from the Galactic plane, $|Z|$. The Galactic latitudes of VY Pyx and \kp~, $+13\fdg6$ and $-25\fdg4$ respectively, together with the Table~\ref{tbl-SUM2} parallaxes, yield Z values of 36.5pc and 77pc, not particularly extreme. Nor are the transverse velocities of VY Pyx and \kp~(Table~\ref{tbl-SUM2}) indicative of halo or thick disk membership. The metallicity of \kp, [Fe/H]=0.0 \citep{Luc89} is far from the prototypical [Fe/H]=-1. In contrast our newly measured metallicity for VY Pyx ([Fe/H]=-0.01, Table~\ref{tbl-AP}) is the same 
as that measured by \cite{Maa07}  for the prototype BL Her (though their value for VY Pyx
is -0.4).

Figure~\ref{Fig6} shows  an M$_K$ - log\,P relation. The slope (-2.38 $\pm$ 0.04) was derived by \cite{Sol06} from RRLs 
in globular clusters. This is essentially the same as the slope found by \cite{Mat06} for
CP2s in globular clusters (-2.41 $\pm$ 0.05). The zero-point was fixed by our RRLs which are shown. 
This relation passes within 0.01mag of our
absolute magnitude of \kp, which is also plotted, although this star was not used in deriving the zero-point. 
This suggests that either \kp~is not a peculiar W Vir star, or is one of the few that lie near the PL 
relation. It is very desirable to clear up this matter. If it can be used as a normal CP2, it
would add significantly to the distance scale calibration.

VY Pyx at M$_{K}$, log\,P = -0.26, 0.0934 lies +1.19$\pm$ 0.08 mags below the regression line of Figure~\ref{Fig6}.
A weighted average of HIP 97 and HIP07 parallaxes ($\pi_{abs} =5.37 \pm 0.38$) gives M$_{K} =-0.68 \pm 0.16$
which is +0.78 magnitudes below the regression line. It is not clear whether this star indicates that a wide range of absolute magnitudes is possible for short period CP2s, or whether it is a rare anomaly. Detailed
studies of CP2s with periods near one day in the Magellanic Clouds may answer this question. 

\section{Summary} \label{SUMM}
\begin{enumerate}
\item {\it HST} astrometry has now yielded absolute trigonometric parallaxes  for 5 RRL variables and two CP2 with an average $\sigma_\pi = 0.18 $ mas, or $\sigma_\pi/\pi = 5.4$\%. These parallaxes, along with precision photometry culled from the literature, Lutz-Kelker-Hanson bias corrections,  and reddening corrections derived from both the literature and/or our ancillary spectrophotometry, provide absolute magnitudes with which to extract zero-points for a Period-Luminosity Relation and an M$_V$ - [Fe/H] relation. The restricted ranges of both log\,P and [Fe/H] preclude solving for slopes. Adopting previously determined slopes, our zero-point errors are now at or below 0.03 magnitudes in the K bandpass and 0.05 in the V bandpass.

\item To obtain these parallaxes, no RRL or CP2 required the addition of a perturbation orbit in the modeling.

\item The CP2 \kp~ (log\,P = 0.96) lies within 0.01 magnitude of the value predicted by an extrapolation of an RRL K$_{s}$ - log\,P
relation based on a slope derived from globular clusters and  our  parallax
zero-point. This star could be an important distance scale calibrator, if the
uncertainty regarding its nature (normal or peculiar CP2) can be resolved. This situation appears to support the assertion that RRL and CP2 together can establish a single PLR \citep{Mat06,Maj10}.

\item  Our absolute magnitude of the CP2 star VY Pyx (log\,P = 0.093) falls well below
a K$_s$ - log\,P relation for RRLs based on our zero-point. This result is not currently understood but we
see no reason to question the accuracy of our parallax.

\item We apply our V and K calibrations to selected galactic globular clusters. We obtain K-band PLR and M$_V$ - [Fe/H] distance moduli that agree within the errors for all clusters.  Ages obtained from stellar evolution models range 10.8 -- 13.1 Gy.

\item Based on the K$_{s}$ -[Fe/H] - log\,P relation of \cite{Bor09}
and with our zero-point calibration, we derive an LMC distance modulus of 
(m-M)$_{0} = 18.55 \pm 0.05$.  This result agrees within the errors with
that derived from classical Cepheids, calibrated by \HST~parallaxes
\citep{Ben07} and uncorrected for metallicity differences between the
Galactic calibrators and the LMC Cepheids.

\end{enumerate}

\acknowledgments

Support for this work was provided by NASA through grants GO-11211 and GO-11789  from the Space Telescope Science Institute, which is operated
by AURA, Inc., under NASA contract NAS5-26555. This paper uses observations made at the South African Astronomical Observatory (SAAO), and observations obtained with the Apache Point Observatory 3.5-meter telescope, which is owned and operated by the Astrophysical Research Consortium. This paper uses observations made at the Cerro Tololo Observatory 4m telescope. Cerro Tololo is also operated by AURA. TEH thanks the SMARTS crew for obtaining photometry of the VY Pyx and UV Oct fields after three weathered-out failed attempts by TEH. This publication makes use of data products from the Two Micron All Sky Survey, which is a joint project of the University of Massachusetts and the Infrared Processing and Analysis Center/California Institute of Technology, funded by NASA and the NSF. This research has made use of the SIMBAD database, operated at CDS, Strasbourg, France; the NASA/IPAC Extragalactic Database (NED) which is operated by JPL, California Institute of Technology, under contract with the NASA;  and NASA's Astrophysics Data System Abstract Service.  This material is based in part upon work by TGB while serving at the National Science Foundation. Any opinions, findings, and conclusions or recommendations expressed in this material are those of the authors and do not necessarily reflect the views of the National Science Foundation. This work is supported by an STFC Rolling Grant (L.F.). O.K. is a Royal Swedish
Academy of Sciences Research Fellow supported by grants from the Knut and
Alice Wallenberg Foundation and the Swedish Research Council. D.S. acknowledges
support received from the Deutsche Forschungsgemeinschaft (DFG) Research
Grant RE1664/7-1. KK, LF and NN are supported by Austrian FWF grants T359 and P19962. Feast and Menzies thank the National
Research Foundation (NRF) of South Africa for financial support.
GFB thanks Debbie Winegarten, whose able assistance with other matters freed me to devote necessary time to this project. Finally we thank an anonymous referee for careful reading and extensive constructive criticism which materially improved the presentation and discussion.
\clearpage


\bibliography{/Active/myMaster}

\clearpage
\begin{deluxetable}{l l l l l l l l}
\tablewidth{6in}
\tablecaption{Target Properties \label{tbl-AP}} 
\tablehead{\colhead{ID}&
\colhead{log P} &
\colhead{T$_0$}&
\colhead{$\langle$V$\rangle$} &
\colhead{$\langle$K$_S\rangle$\tablenotemark{a}} &
\colhead{[Fe/H]\tablenotemark{b}} &
\colhead{A$_V$} &
\colhead{A$_K$} 
}
\startdata
RZ Cep (c)\tablenotemark{c}&-0.51052&54793.0050&9.47&8.11&-1.77$\pm$0.2&0.78\\
XZ Cyg (ab)\tablenotemark{d}&-0.33107&54395.1020&9.68&8.72&-1.44 0.2&0.30&0.04\\
SU Dra (ab)\tablenotemark{e}&-0.18018&54733.1510&9.78&8.62&-1.80 0.2&0.03&0.00\\
RR Lyr (ab)\tablenotemark{f}&-0.24655&50749.2380&7.76&6.49&-1.41 0.13&0.13&0.01\\
UV Oct (ab)\tablenotemark{g}&-0.26552&53570.4141&9.50&8.30&-1.47 0.11&0.28&0.03\\
VY Pyx (BLHer)\tablenotemark{h}&0.09340&54406.4072&7.30&5.72&-0.01 0.15&0.15&0.02\\
\kp~(WVir)\tablenotemark{i}&0.95815&54705.9320&4.35&2.78&0.0 0.13&0.05&0.0\\
\enddata
\tablenotetext{a}{K$_S$ from \cite{Fea08}, except where noted.}
\tablenotetext{b}{[Fe/H] on \cite{Zin84} scale.}
\tablenotetext{c}{RZ Cep: T$_0$, Smith for this paper; [Fe/H],  A$_V$, \cite{Fer98a}}
\tablenotetext{d}{XZ Cyg: \cite{Fer98a}, \cite{Fea08}; logP, T$_0$ from \cite{LaC04}}
\tablenotetext{e}{SU Dra:  \cite{Fer98a}, \cite{Fea08}}
\tablenotetext{f}{RR Lyr:  \cite{Fer98a}, \cite{Fea08}; [Fe/H], A$_V$, \cite{Kol10}}
\tablenotetext{g}{UV Oct:  \cite{Fer98a}, \cite{Fea08}; logP, T$_0$ from FGS photometry; [Fe/H] derived as for RR Lyr for this paper}
\tablenotetext{h}{VY Pyx:  \cite{Fea08}; $\langle$K$_S\rangle$  from Laney for this paper; T$_0$ from FGS photometry; [Fe/H] derived as for RR Lyr for this paper}
\tablenotetext{i}{\kp : \cite{Fea08}; T$_0$ from FGS photometry; [Fe/H], \cite{Luc89}}
\end{deluxetable}

\begin{deluxetable}{lllllrrrrr}
\tablewidth{0in}
\tablecaption{Log of Observations,  Apparent Magnitude, Estimated B-V, and Pulsational Phase\label{tbl-LOO}}
\tablehead{\colhead{Set}&
\colhead{mJD}&\colhead{V}&\colhead{B-V\tablenotemark{a}} & \colhead{Phase}
&\colhead{Set}&
\colhead{mJD}&\colhead{V}&\colhead{B-V\tablenotemark{a}} & \colhead{Phase}
}
\startdata
RZ Cep&&&&&XZ Cyg&&&&\\
1&54287.60885&9.759&0.53&0.6121&1&54292.22174&10.092&0.40&0.5050\\
1&54287.62167&9.719&0.52&0.6536&1&54292.2351&10.129&0.40&0.5106\\
1&54287.63148&9.704&0.52&0.6854&1&54292.24162&10.115&0.41&0.5337\\
1&54287.6403&9.683&0.51&0.7140&1&54292.24659&10.126&0.41&0.5476\\
1&54287.65023&9.643&0.50&0.7462&2&54322.03022&9.976&0.38&0.3911\\
1&54287.65756&9.599&0.49&0.7699&2&54322.03628&9.996&0.39&0.4041\\
2&54342.38356&9.270&0.41&0.0718&2&54322.04291&10.020&0.39&0.4183\\
2&54342.39431&9.309&0.41&0.1066&2&54322.04787&10.036&0.39&0.4290\\
2&54342.40291&9.337&0.42&0.1345&2&54322.05285&10.051&0.39&0.4396\\
2&54342.41148&9.363&0.42&0.1623&2&54322.05709&10.064&0.39&0.4487\\
2&54342.41535&9.375&0.42&0.1748&3&54348.96281&9.430&0.24&0.1136\\
2&54342.4233&9.402&0.43&0.2005&3&54348.96887&9.463&0.25&0.1266\\
2&54342.43593&9.440&0.44&0.2414&3&54348.97549&9.499&0.26&0.1408\\
3&54394.32964&9.596&0.48&0.3673&3&54348.98044&9.525&0.27&0.1514\\
3&54394.34037&9.635&0.49&0.4020&3&54348.98544&9.553&0.28&0.1621\\
3&54394.34896&9.667&0.50&0.4299&3&54348.98968&9.575&0.29&0.1712\\
3&54394.35756&9.696&0.50&0.4577&4&54395.08432&9.603&0.19&0.9621\\
3&54394.36141&9.706&0.51&0.4702&4&54395.09038&9.344&0.18&0.9751\\
3&54394.36936&9.721&0.51&0.4960&4&54395.097&9.105&0.17&0.9893\\
3&54394.382&9.732&0.52&0.5369&4&54395.10194&9.014&0.16&0.9999\\
4&54448.08932&9.731&0.52&0.5384&4&54395.10694&8.975&0.16&0.0106\\
4&54448.09977&9.725&0.53&0.5723&4&54395.11119&8.974&0.16&0.0197\\
4&54448.1081&9.716&0.53&0.5993&5&54489.86693&9.334&0.22&0.1020\\
4&54448.11644&9.705&0.53&0.6263&5&54489.87301&9.373&0.23&0.1151\\
5&54484.14804&9.588&0.47&0.3619&5&54489.87963&9.416&0.25&0.1292\\
5&54484.1585&9.628&0.49&0.3958&5&54489.88462&9.448&0.26&0.1399\\
5&54484.16683&9.659&0.49&0.4228&5&54489.88956&9.480&0.26&0.1505\\
5&54484.17516&9.686&0.50&0.4498&5&54489.89382&9.506&0.27&0.1597\\
6&54580.05927&9.309&0.41&0.0961&6&54516.76767&10.190&0.39&0.7562\\
6&54580.06972&9.350&0.42&0.1300&6&54516.77373&10.191&0.39&0.7692\\
6&54580.07806&9.381&0.42&0.1570&6&54516.78035&10.189&0.38&0.7834\\
6&54580.08639&9.408&0.42&0.1840&6&54516.78532&10.190&0.37&0.7941\\
7&54622.23102&9.602&0.51&0.7246&6&54516.79029&10.193&0.36&0.8047\\
7&54622.24147&9.508&0.50&0.7585&6&54516.79454&10.196&0.36&0.8138\\
7&54622.2498&9.399&0.49&0.7855&7&54578.98769&9.417&0.22&0.1074\\
7&54622.25814&9.285&0.47&0.8125&7&54578.99372&9.444&0.23&0.1203\\
8&54677.10959&9.724&0.52&0.5208&7&54579.00032&9.476&0.25&0.1344\\
8&54677.12003&9.721&0.52&0.5546&7&54579.00529&9.500&0.26&0.1451\\
8&54677.12838&9.715&0.53&0.5817&7&54579.01028&9.524&0.27&0.1558\\
8&54677.1367&9.704&0.53&0.6086&7&54579.01453&9.543&0.27&0.1649\\
9&54730.0075&9.225&0.43&0.9000&8&54623.24464&9.588&0.21&0.9597\\
9&54730.01795&9.206&0.42&0.9339&8&54623.25071&9.472&0.19&0.9727\\
9&54730.02628&9.204&0.41&0.9609&8&54623.25734&9.341&0.18&0.9869\\
9&54730.03462&9.217&0.40&0.9879&8&54623.26228&9.254&0.17&0.9975\\
10&54778.54648&9.388&0.42&0.1571&8&54623.26728&9.196&0.17&0.0082\\
10&54778.55722&9.421&0.43&0.1919&8&54623.27152&9.167&0.16&0.0173\\
10&54778.56581&9.447&0.43&0.2198&9&54702.39156&10.112&0.41&0.5889\\
10&54778.5744&9.478&0.44&0.2476&9&54702.3966&10.118&0.41&0.5997\\
10&54778.57826&9.490&0.44&0.2601&9&54702.40324&10.130&0.42&0.6140\\
10&54778.58622&9.516&0.45&0.2859&9&54702.40819&10.138&0.42&0.6246\\
10&54778.59884&9.561&0.46&0.3268&9&54702.41318&10.145&0.42&0.6353\\
11&54837.09566&9.214&0.46&0.8454&9&54702.41742&10.151&0.42&0.6444\\
11&54837.10639&9.212&0.44&0.8802&10&54781.03128&9.368&0.23&0.1311\\
11&54837.11498&9.219&0.43&0.9080&10&54781.03465&9.390&0.24&0.1383\\
11&54837.12355&9.202&0.42&0.9358&10&54781.03947&9.422&0.25&0.1486\\
11&54837.12742&9.195&0.41&0.9483&10&54781.04615&9.466&0.26&0.1630\\
11&54837.13537&9.194&0.41&0.9741&10&54781.05123&9.497&0.27&0.1738\\
11&54837.1473&9.261&0.40&0.0127&10&54781.05826&9.540&0.28&0.1889\\
12&54959.36343&9.204&0.41&0.9698&11&54954.92647&10.147&0.37&0.8266\\
12&54959.37416&9.193&0.40&0.0046&11&54954.92878&10.147&0.36&0.8316\\
12&54959.38274&9.203&0.41&0.0324&11&54954.93449&10.153&0.35&0.8438\\
12&54959.39133&9.225&0.41&0.0602&11&54954.94096&10.155&0.34&0.8577\\
12&54959.3952&9.238&0.41&0.0728&11&54954.94594&10.152&0.33&0.8684\\
12&54959.40315&9.268&0.41&0.0985&11&54954.9541&10.134&0.31&0.8858\\
12&54959.41578&9.315&0.42&0.1394&12&54997.46322&9.352&0.19&0.9922\\
13&55168.48128&9.685&0.51&0.4721&12&54997.46659&9.325&0.19&0.9994\\
13&55168.492&9.713&0.52&0.5068&12&54997.47241&9.298&0.18&0.0119\\
13&55168.50059&9.728&0.52&0.5347&12&54997.47797&9.287&0.17&0.0238\\
13&55168.50918&9.734&0.52&0.5625&12&54997.48304&9.289&0.16&0.0347\\
13&55168.51304&9.734&0.53&0.5750&12&54997.49095&9.313&0.16&0.0517\\
13&55168.521&9.729&0.53&0.6008&13&55176.44902&10.127&0.41&0.5980\\
13&55168.53225&9.712&0.53&0.6372&13&55176.45135&10.126&0.41&0.6031\\
&&&&&13&55176.45705&10.127&0.41&0.6153\\
&&&&&13&55176.46355&10.127&0.41&0.6292\\
&&&&&13&55176.46854&10.128&0.42&0.6399\\
&&&&&13&55176.4763&10.123&0.42&0.6565\\
&&&&&14&55341.16571&10.141&0.41&0.6220\\
&&&&&14&55341.16804&10.142&0.41&0.6270\\
&&&&&14&55341.17571&10.152&0.41&0.6435\\
&&&&&14&55341.18219&10.157&0.42&0.6574\\
&&&&&14&55341.18713&10.158&0.42&0.6680\\
&&&&&14&55341.193&10.162&0.42&0.6805\\
SU Dra& & & &&RR Lyr& & & &\\
2&54396.06873&10.195&0.38&0.5930&1&49984.25103&7.68&0.38&0.417\\
2&54396.08065&10.195&0.38&0.6111&1&49984.25581&7.68&0.38&0.425\\
2&54396.09115&10.191&0.39&0.6263&1&49984.26076&7.69&0.38&0.434\\
2&54396.0953&10.190&0.39&0.6323&1&49984.26525&7.69&0.38&0.442\\
3&54426.18972&9.794&0.28&0.2019&1&49984.27176&7.7&0.38&0.453\\
3&54426.20164&9.826&0.30&0.2201&1&49984.27638&7.71&0.38&0.461\\
3&54426.21214&9.851&0.31&0.2353&4&50047.04674&7.36&0.30&0.200\\
3&54426.21628&9.859&0.31&0.2413&4&50047.05186&7.37&0.31&0.209\\
4&54478.92392&9.472&0.14&0.0514&4&50047.05669&7.39&0.32&0.218\\
4&54478.93418&9.504&0.15&0.0665&4&50047.0617&7.4&0.32&0.227\\
4&54478.94468&9.540&0.16&0.0832&4&50047.06661&7.41&0.33&0.235\\
4&54478.94877&9.554&0.17&0.0892&4&50047.07193&7.43&0.34&0.245\\
4&54478.95101&9.562&0.17&0.0923&5&50172.97726&7.63&0.39&0.366\\
4&54478.96269&9.602&0.19&0.1104&5&50172.98699&7.64&0.39&0.383\\
4&54478.97122&9.626&0.21&0.1225&5&50172.9915&7.65&0.39&0.391\\
4&54478.97532&9.638&0.21&0.1286&5&50172.99797&7.66&0.38&0.402\\
5&54492.91603&9.852&0.31&0.2380&5&50173.00264&7.67&0.38&0.411\\
5&54492.92925&9.886&0.32&0.2576&6&50186.85366&7.91&0.42&0.847\\
5&54492.94056&9.908&0.33&0.2758&6&50186.85845&7.91&0.41&0.855\\
5&54492.94468&9.916&0.33&0.2819&6&50186.86339&7.93&0.40&0.864\\
6&54532.48479&9.701&0.24&0.1530&6&50186.86788&7.93&0.39&0.872\\
6&54532.49495&9.729&0.25&0.1667&6&50186.87441&7.94&0.37&0.883\\
6&54532.50539&9.758&0.27&0.1833&6&50186.879&7.95&0.36&0.891\\
6&54532.50948&9.770&0.27&0.1894&7&50201.05711&7.94&0.34&0.904\\
6&54532.51167&9.775&0.28&0.1939&7&50201.06191&7.93&0.33&0.913\\
6&54532.52188&9.804&0.29&0.2106&7&50201.06684&7.93&0.31&0.921\\
6&54532.53042&9.825&0.30&0.2242&7&50201.07133&7.92&0.30&0.929\\
6&54532.53456&9.836&0.30&0.2303&7&50201.07787&7.89&0.28&0.941\\
7&54586.25231&10.199&0.37&0.5666&7&50201.08245&7.86&0.27&0.949\\
7&54586.26252&10.207&0.38&0.5833&8&50228.80167&7.9&0.42&0.851\\
7&54586.27297&10.214&0.38&0.5984&8&50228.80661&7.91&0.41&0.860\\
7&54586.27706&10.212&0.38&0.6045&8&50228.81119&7.92&0.40&0.868\\
7&54586.27925&10.210&0.38&0.6075&8&50228.81616&7.93&0.38&0.876\\
7&54586.28946&10.206&0.39&0.6257&8&50228.8208&7.94&0.37&0.885\\
7&54586.298&10.203&0.39&0.6378&8&50228.82721&7.94&0.35&0.896\\
7&54586.30214&10.201&0.39&0.6439&9&50562.87498&7.51&0.32&0.220\\
8&54639.00762&10.096&0.35&0.4494&9&50562.88171&7.52&0.33&0.232\\
8&54639.01944&10.113&0.35&0.4660&9&50562.88664&7.53&0.34&0.241\\
8&54639.02994&10.126&0.36&0.4827&9&50562.89113&7.54&0.34&0.249\\
8&54639.03405&10.130&0.36&0.4887&9&50562.89767&7.55&0.35&0.260\\
9&54733.1287&9.477&0.20&0.9667&9&50562.90226&7.56&0.36&0.268\\
9&54733.1406&9.477&0.20&0.9667&10&50567.58263&7.77&0.39&0.525\\
9&54733.1511&9.388&0.17&0.9849&10&50567.58936&7.77&0.39&0.537\\
9&54733.15524&9.388&0.17&0.9849&10&50567.59429&7.77&0.40&0.546\\
10&54833.63013&9.660&0.23&0.1446&10&50567.59878&7.78&0.40&0.554\\
10&54833.64205&9.697&0.25&0.1627&10&50567.60532&7.78&0.40&0.565\\
10&54833.65253&9.726&0.26&0.1794&10&50567.60991&7.78&0.41&0.573\\
10&54833.65667&9.738&0.27&0.1854&11&50745.19083&7.94&0.41&0.860\\
12&55130.18997&9.761&0.28&0.1935&11&50745.19752&7.95&0.39&0.872\\
12&55130.19992&9.791&0.29&0.2086&11&50745.20249&7.95&0.38&0.881\\
12&55130.2095&9.819&0.30&0.2238&11&50745.20698&7.96&0.37&0.889\\
12&55130.21683&9.837&0.31&0.2344&11&50745.2135&7.95&0.35&0.900\\
13&55143.86819&10.080&0.28&0.9046&11&50745.2181&7.94&0.33&0.908\\
13&55143.87203&10.005&0.28&0.9107&12&50749.22225&7.36&0.23&0.972\\
13&55143.87913&9.831&0.26&0.9213&12&50749.22894&7.26&0.22&0.984\\
13&55143.88887&9.711&0.24&0.9364&12&50749.2339&7.19&0.21&0.993\\
13&55143.89302&9.673&0.23&0.9425&12&50749.2384&7.13&0.17&0.001\\
13&55143.90014&9.585&0.22&0.9531&12&50749.24358&7.09&0.16&0.010\\
13&55143.90985&9.462&0.19&0.9682&12&50749.24821&7.06&0.15&0.018\\
13&55143.91834&9.395&0.18&0.9803&13&54781.1015&7.81&0.22&0.978\\
15&55149.26255&9.522&0.15&0.0737&13&54781.10757&7.73&0.21&0.989\\
15&55149.2725&9.556&0.17&0.0903&13&54781.11395&7.66&0.20&1.000\\
15&55149.28207&9.587&0.19&0.1040&13&54781.12127&7.58&0.16&0.013\\
15&55149.28941&9.610&0.19&0.1130&13&54781.12775&7.51&0.15&0.024\\
16&55317.09729&9.793&0.29&0.2067&14&54781.16808&7.4&0.17&0.095\\
16&55317.10911&9.824&0.30&0.2248&14&54781.17414&7.42&0.18&0.106\\
16&55317.11961&9.850&0.31&0.2415&14&54781.18052&7.44&0.19&0.117\\
16&55317.12375&9.859&0.31&0.2476&14&54781.18786&7.47&0.21&0.130\\
&&&&&14&54781.19433&7.49&0.22&0.142\\
&&&&&15&54781.36779&7.84&0.38&0.448\\
&&&&&15&54781.37385&7.85&0.38&0.458\\
&&&&&15&54781.38023&7.86&0.38&0.470\\
&&&&&15&54781.38756&7.86&0.38&0.483\\
&&&&&15&54781.39404&7.87&0.38&0.494\\
&&&&&16&54782.10012&7.91&0.48&0.740\\
&&&&&16&54782.10618&7.91&0.48&0.750\\
&&&&&16&54782.11256&7.91&0.48&0.762\\
&&&&&16&54782.11988&7.91&0.48&0.775\\
&&&&&16&54782.12637&7.91&0.47&0.786\\
&&&&&17&54784.16409&7.78&0.39&0.381\\
&&&&&17&54784.17015&7.79&0.39&0.392\\
&&&&&17&54784.17653&7.79&0.38&0.403\\
&&&&&17&54784.18385&7.8&0.38&0.416\\
&&&&&17&54784.19034&7.81&0.38&0.427\\
UV Oct& & & &&VY Pyx& & & &\\
1&54280.88854&9.663&0.38&0.3749&1&54391.79701&7.284&0.55&0.2171\\
1&54280.89363&9.677&0.38&0.3889&1&54391.80337&7.290&0.55&0.2223\\
1&54280.90203&9.696&0.38&0.3983&1&54391.80916&7.289&0.56&0.2269\\
1&54280.90706&9.708&0.38&0.4138&1&54391.81659&7.293&0.56&0.2329\\
2&54321.13128&9.695&0.38&0.4231&1&54391.82215&7.296&0.56&0.2374\\
2&54321.13642&9.697&0.44&0.6712&2&54399.99065&7.227&0.54&0.8252\\
2&54321.14453&9.704&0.44&0.6806&2&54399.997&7.223&0.54&0.8303\\
2&54321.1551&9.717&0.44&0.6956&2&54400.00278&7.222&0.54&0.8350\\
2&54321.16021&9.724&0.44&0.7151&2&54400.01022&7.215&0.53&0.8410\\
3&54371.95272&9.421&0.44&0.7245&2&54400.01578&7.212&0.53&0.8454\\
3&54371.95787&9.432&0.30&0.1851&3&54406.38762&7.194&0.51&0.9842\\
3&54371.96595&9.451&0.31&0.1945&3&54406.39397&7.194&0.51&0.9893\\
3&54371.97653&9.476&0.32&0.2094&3&54406.39975&7.194&0.50&0.9940\\
3&54371.98163&9.488&0.33&0.2289&3&54406.40719&7.194&0.50&0.0000\\
4&54376.00938&9.648&0.34&0.2383&3&54406.41275&7.197&0.50&0.0045\\
4&54376.01451&9.662&0.44&0.6945&4&54414.18113&7.313&0.57&0.2695\\
4&54376.0226&9.689&0.44&0.7040&4&54414.18749&7.315&0.57&0.2747\\
4&54376.03318&9.727&0.44&0.7189&4&54414.19328&7.318&0.57&0.2793\\
4&54376.0383&9.745&0.43&0.7384&4&54414.20071&7.321&0.57&0.2853\\
5&54522.69065&9.172&0.43&0.7479&4&54414.20626&7.323&0.57&0.2898\\
5&54522.69579&9.185&0.16&0.0250&5&54421.04014&7.242&0.55&0.8012\\
5&54522.70389&9.208&0.16&0.0344&5&54421.04649&7.237&0.54&0.8063\\
5&54522.71447&9.233&0.17&0.0494&5&54421.05229&7.235&0.54&0.8110\\
5&54522.71955&9.243&0.19&0.0689&5&54421.05972&7.228&0.54&0.8170\\
6&54551.36091&9.377&0.20&0.0782&5&54421.06527&7.227&0.54&0.8215\\
6&54551.36606&9.243&0.34&0.8636&6&54427.69953&7.266&0.54&0.1719\\
6&54551.37417&9.033&0.33&0.8731&6&54427.70588&7.268&0.54&0.1770\\
6&54551.38473&8.868&0.31&0.8881&6&54427.71167&7.269&0.54&0.1817\\
6&54551.38984&8.834&0.29&0.9075&6&54427.7191&7.273&0.54&0.1877\\
7&54569.72058&9.757&0.28&0.9169&6&54427.72465&7.276&0.54&0.1922\\
7&54569.72574&9.762&0.44&0.7001&7&54439.55301&7.300&0.57&0.7315\\
7&54569.73384&9.778&0.44&0.7096&7&54439.55938&7.297&0.57&0.7367\\
7&54569.74442&9.797&0.44&0.7245&7&54439.56516&7.293&0.56&0.7413\\
7&54569.74951&9.806&0.43&0.7440&7&54439.57259&7.286&0.56&0.7473\\
8&54601.98409&9.356&0.43&0.7534&7&54439.57815&7.284&0.56&0.7518\\
8&54601.98925&9.368&0.28&0.1610&8&54458.27282&7.213&0.54&0.8287\\
8&54601.99733&9.390&0.29&0.1706&8&54458.2792&7.209&0.54&0.8339\\
8&54602.00791&9.418&0.30&0.1854&8&54458.28498&7.205&0.54&0.8385\\
8&54602.01302&9.430&0.31&0.2049&8&54458.29241&7.202&0.53&0.8445\\
9&54660.78087&9.629&0.32&0.2144&8&54458.29796&7.200&0.53&0.8490\\
9&54660.78602&9.629&0.40&0.5222&9&54466.59728&7.424&0.60&0.5423\\
9&54660.79411&9.630&0.40&0.5317&9&54466.60366&7.423&0.60&0.5474\\
9&54660.80468&9.635&0.41&0.5466&9&54466.60944&7.424&0.60&0.5521\\
9&54660.80979&9.639&0.41&0.5661&9&54466.61686&7.423&0.60&0.5581\\
10&54704.02235&9.446&0.42&0.5755&9&54466.62242&7.421&0.60&0.5625\\
10&54704.02749&9.458&0.32&0.2153&10&54471.59201&7.416&0.60&0.5704\\
10&54704.03559&9.479&0.33&0.2248&10&54471.59838&7.413&0.60&0.5756\\
10&54704.04617&9.508&0.34&0.2397&10&54471.60417&7.413&0.60&0.5802\\
10&54704.05126&9.523&0.35&0.2592&10&54471.6116&7.410&0.60&0.5862\\
11&54900.0444&9.714&0.35&0.2686&10&54471.61716&7.408&0.60&0.5907\\
11&54900.04955&9.717&0.39&0.4797&11&54482.71326&7.412&0.60&0.5395\\
11&54900.05764&9.719&0.39&0.4892&11&54482.71963&7.412&0.60&0.5447\\
11&54900.06822&9.724&0.40&0.5041&11&54482.72542&7.411&0.60&0.5493\\
11&54900.07331&9.727&0.40&0.5236&11&54482.73285&7.411&0.60&0.5553\\
12&54908.69757&9.660&0.40&0.5330&11&54482.7384&7.406&0.60&0.5598\\
12&54908.70125&9.662&0.38&0.4273&12&54491.30479&7.406&0.60&0.4685\\
12&54908.70933&9.668&0.38&0.4341&12&54491.31115&7.405&0.60&0.4736\\
12&54908.71991&9.670&0.39&0.4490&12&54491.31693&7.408&0.61&0.4782\\
12&54908.725&9.671&0.39&0.4685&12&54491.32436&7.409&0.61&0.4842\\
13&55075.29118&9.626&0.39&0.4779&12&54491.32993&7.411&0.61&0.4887\\
13&55075.29632&9.628&0.39&0.4557&13&54499.62819&7.271&0.54&0.1811\\
13&55075.30442&9.631&0.39&0.4652&13&54499.63456&7.272&0.54&0.1863\\
13&55075.315&9.637&0.39&0.4801&13&54499.64035&7.273&0.54&0.1909\\
13&55075.32008&9.640&0.40&0.4996&13&54499.64778&7.275&0.55&0.1969\\
14&55087.85755&9.681&0.40&0.5090&13&54499.65333&7.277&0.55&0.2014\\
14&55087.86269&9.682&0.43&0.6153&14&54526.47303&7.217&0.54&0.8310\\
14&55087.87078&9.696&0.43&0.6247&14&54526.47941&7.215&0.54&0.8362\\
14&55087.88135&9.725&0.43&0.6396&14&54526.48519&7.212&0.53&0.8408\\
14&55087.88646&9.740&0.44&0.6591&14&54526.49262&7.208&0.53&0.8468\\
15&55148.83507&9.040&0.44&0.6685&14&54526.49818&7.207&0.53&0.8513\\
15&55148.84021&9.066&0.18&0.9955&15&54532.39944&7.395&0.60&0.6106\\
15&55148.8483&9.110&0.15&0.0050&15&54532.40581&7.392&0.59&0.6157\\
15&55148.85888&9.161&0.16&0.0199&15&54532.4116&7.384&0.59&0.6204\\
15&55148.86398&9.183&0.17&0.0394&15&54532.41903&7.381&0.59&0.6264\\
&&&&&15&54532.42458&7.380&0.59&0.6309\\
&&&&&16&54799.06059&7.353&0.58&0.6682\\
&&&&&16&54799.06696&7.351&0.58&0.6733\\
&&&&&16&54799.07274&7.347&0.58&0.6780\\
&&&&&16&54799.08017&7.340&0.58&0.6840\\
&&&&&16&54799.08573&7.338&0.58&0.6885\\
&&&&&17&54817.77185&7.279&0.56&0.7585\\
&&&&&17&54817.77822&7.272&0.56&0.7636\\
&&&&&17&54817.784&7.268&0.56&0.7683\\
&&&&&17&54817.79144&7.265&0.55&0.7743\\
&&&&&17&54817.79699&7.260&0.55&0.7788\\
&&&&&18&54860.52883&7.285&0.56&0.2413\\
&&&&&18&54860.53521&7.287&0.56&0.2464\\
&&&&&18&54860.541&7.292&0.56&0.2511\\
&&&&&18&54860.54843&7.297&0.56&0.2571\\
&&&&&18&54860.55398&7.298&0.57&0.2616\\
&&&&&19&55289.58789&7.310&0.57&0.2700\\
&&&&&19&55289.59426&7.309&0.57&0.2752\\
&&&&&19&55289.60003&7.314&0.57&0.2798\\
&&&&&19&55289.60747&7.318&0.57&0.2858\\
&&&&&19&55289.61303&7.318&0.57&0.2903\\
&&&&&20&55301.43772&7.222&0.54&0.8267\\
&&&&&20&55301.44407&7.218&0.54&0.8318\\
&&&&&20&55301.44985&7.215&0.54&0.8365\\
&&&&&20&55301.45728&7.211&0.53&0.8425\\
&&&&&20&55301.46285&7.207&0.53&0.8470\\
&&&&&21&55325.93291&7.409&0.60&0.5816\\
&&&&&21&55325.93927&7.408&0.60&0.5868\\
&&&&&21&55325.94506&7.406&0.60&0.5914\\
&&&&&21&55325.95249&7.404&0.60&0.5974\\
&&&&&21&55325.95804&7.403&0.60&0.6019\\
&&&&&22&55332.25647&7.344&0.58&0.6815\\
&&&&&22&55332.26284&7.341&0.58&0.6866\\
&&&&&22&55332.26862&7.339&0.58&0.6913\\
&&&&&22&55332.27605&7.334&0.58&0.6973\\
&&&&&22&55332.28161&7.331&0.58&0.7018\\
&&&&&23&55351.36845&7.227&0.52&0.0950\\
&&&&&23&55351.37481&7.229&0.52&0.1001\\
&&&&&23&55351.3806&7.230&0.52&0.1048\\
&&&&&23&55351.38803&7.231&0.52&0.1108\\
&&&&&23&55351.39359&7.234&0.52&0.1152\\
\kp&&&&\\
1&54280.81179&4.207&0.60&0.1921\\
1&54280.82106&4.210&0.61&0.1931\\
1&54280.82566&4.212&0.61&0.1936\\
1&54280.83559&4.212&0.61&0.1947\\
2&54321.05596&4.633&0.86&0.6235\\
2&54321.06751&4.634&0.86&0.6248\\
2&54321.07611&4.631&0.86&0.6258\\
2&54321.08404&4.633&0.86&0.6266\\
2&54321.09235&4.633&0.86&0.6276\\
2&54321.10119&4.631&0.86&0.6285\\
3&54373.07462&4.472&0.82&0.3516\\
3&54373.08557&4.474&0.82&0.3528\\
3&54373.09296&4.475&0.83&0.3536\\
3&54373.1042&4.477&0.83&0.3548\\
3&54373.11263&4.480&0.83&0.3558\\
3&54373.12443&4.482&0.83&0.3571\\
4&54519.03565&4.569&0.89&0.4241\\
4&54519.04564&4.571&0.89&0.4252\\
4&54519.05097&4.572&0.89&0.4258\\
4&54519.06171&4.575&0.89&0.4270\\
5&54548.08718&4.662&0.86&0.6231\\
5&54548.09788&4.660&0.86&0.6243\\
5&54548.10277&4.659&0.86&0.6248\\
5&54548.11111&4.660&0.86&0.6258\\
5&54548.11814&4.660&0.86&0.6265\\
5&54548.12516&4.657&0.86&0.6273\\
7&54601.02693&4.613&0.90&0.4526\\
7&54601.03856&4.614&0.90&0.4539\\
7&54601.04464&4.613&0.90&0.4545\\
7&54601.05392&4.612&0.90&0.4556\\
7&54601.06093&4.612&0.90&0.4563\\
7&54601.06792&4.617&0.90&0.4571\\
8&54653.05113&4.192&0.59&0.1813\\
8&54653.06206&4.196&0.59&0.1825\\
8&54653.06906&4.196&0.59&0.1832\\
8&54653.0803&4.196&0.59&0.1845\\
8&54653.08878&4.201&0.59&0.1854\\
8&54653.10058&4.203&0.60&0.1867\\
9&54659.04949&4.219&0.57&0.8418\\
9&54659.06061&4.215&0.57&0.8430\\
9&54659.06693&4.211&0.57&0.8437\\
9&54659.07627&4.205&0.57&0.8447\\
10&54705.88251&3.965&0.42&0.9988\\
10&54705.89347&3.963&0.42&1.0000\\
10&54705.9005&3.963&0.42&0.0008\\
10&54705.9117&3.963&0.42&0.0020\\
10&54705.92017&3.963&0.42&0.0029\\
10&54705.93197&3.961&0.42&0.0042\\
12&54877.21429&4.102&0.54&0.8650\\
12&54877.22524&4.097&0.54&0.8662\\
12&54877.23265&4.094&0.54&0.8670\\
12&54877.24388&4.089&0.54&0.8683\\
12&54877.25233&4.083&0.54&0.8692\\
12&54877.26414&4.078&0.53&0.8705\\
13&54986.28763&4.078&0.53&0.8757\\
13&54986.30008&4.074&0.52&0.8770\\
13&54986.30903&4.069&0.52&0.8780\\
13&54986.32178&4.064&0.52&0.8794\\
13&54986.33166&4.064&0.52&0.8805\\
13&54986.34207&4.061&0.52&0.8816\\
14&55093.89644&4.596&0.75&0.7250\\
14&55093.90741&4.596&0.74&0.7262\\
14&55093.91481&4.596&0.74&0.7270\\
14&55093.92604&4.594&0.74&0.7283\\
14&55093.93447&4.593&0.74&0.7292\\
14&55093.94627&4.594&0.74&0.7305\\
15&55148.88147&4.531&0.67&0.7797\\
15&55148.8887&4.529&0.67&0.7805\\
15&55148.89303&4.527&0.66&0.7810\\
15&55148.89762&4.524&0.66&0.7815\\
15&55148.90877&4.523&0.66&0.7827\\
15&55148.9153&4.523&0.66&0.7834\\
15&55148.92083&4.522&0.66&0.7840\\
15&55148.92597&4.521&0.66&0.7846\\
\enddata
\tablenotetext{a}{B-V estimated from phased light curve.}
\end{deluxetable}

\begin{deluxetable}{l l l l l l l l l}
\tablewidth{0in}
\tablecaption{Reference Stars: Visible and Near-IR Photometry \label{tbl-IR}}
\tablehead{\colhead{FGS ID}&
\colhead{PPMXL ID}&
\colhead{V} &
\colhead{B-V} &
\colhead{K\tablenotemark{a}} &
\colhead{J-K} &
\colhead{V-K} }
\startdata
XZ Cyg&&&&&&\\
2&178627907000819000&15.24&0.79$\pm$0.11&13.34&0.45&1.90\\
3&178627800678871000&15.00&0.72 0.1&13.30&0.37&1.70\\
4&178627507943789000&12.55&1.26 0.03&9.51&0.84&3.04\\
5&178626897495064000&13.16&0.50 0.04&11.96&0.30&1.20\\
6&178627902818755000&12.46&0.65 0.03&11.03&0.37&1.43\\
UV Oct&&&&&&\\
9&6050827457780530000&15.92&0.68 0.04&14.25&0.41&1.67\\
10&6050827543215700000&15.35&0.94 0.05&13.01&0.57&2.35\\
11&6050827517824970000&14.91&0.63 0.04&13.24&0.40&1.67\\
12&6050827495380950000&13.12&1.35 0.03&9.82&0.87&3.30\\
13&6050827141173300000&14.45&0.63 0.04&12.74&0.38&1.71\\
RZ Cep&&&&&&\\
17&236149611067714000& 14.90&1.09 0.07&12.26&0.56&2.64\\
18&236150055842444000& 16.30&1.29 0.15&13.04&0.60&3.26\\
19&236150360634453000& 16.16&0.96 0.15&13.32&0.75&2.84\\
20&236150404085559000& 15.31&1.25 0.1&12.52&0.69&2.79\\
21&236149242851448000& 14.06&1.00 0.06&11.74&0.50&2.32\\
22&236150379961202000& 12.52&0.62 0.03&11.01&0.29&1.51\\
SU Dra&&&&&&\\
24&910626741647084000& 16.26&0.70 0.12&14.51&0.39&1.76\\
25&910625310986521000& 13.10&0.47 0.04&11.80&0.32&1.30\\
26&910625114428228000& 14.60&1.17 0.06&12.14&0.59&2.46\\
27&910626124371472000& 14.36&0.64 0.05&12.66&0.41&1.70\\
28&910627013779397000& 15.19&0.54 0.09&13.75&0.36&1.44\\
\kp&&&&&&\\
31&6417195936484110000&12.65&1.17 0.03&9.98&0.68&2.67\\
32&6417195914995010000&14.24&0.70 0.05&12.56&0.42&1.68\\
33&6417196083800320000&14.55&0.93 0.04&12.39&0.49&2.17\\
34&6417197688468440000&15.94&0.95 0.06&13.72&0.51&2.22\\
35&6417383894358280000&14.84&0.85 0.04&12.96&0.48&1.88\\
36&6417383768816120000&15.31&0.88 0.05&13.30&0.53&2.01\\
37&6417197455800920000&15.7&0.89 0.1&13.74&0.45&1.96\\
VY Pyx&&&&&&\\
39&1264343482\tablenotemark{b}& 12.45&1.34 0.03&9.08&0.89&3.37\\
40&2735010192495240000& 15.31&0.55 0.09&13.77&0.32&1.54\\
41&1264343537\tablenotemark{b}& 15.30&0.93 0.09&13.08&0.53&2.22\\
42&2735033696661810000& 14.41&0.62 0.05&12.71&0.39&1.70\\
43&2735057261695820000& 16.15&0.52 0.15&14.66&0.29&1.49\\

\enddata
\tablenotetext{a}{J, K from 2MASS catalog}
\tablenotetext{b}{ID from 2MASS catalog}
\end{deluxetable}

\begin{deluxetable}{l l l l l r l }
\tablewidth{0in}
\tablecaption{Astrometric Reference Star 
Spectrophotometric Parallaxes \label{tbl-SPP}}
\tablehead{
\colhead{ID}& \colhead{V} &\colhead{Sp. T.}&
 \colhead{M$_V$} & \colhead{A$_V$} &\colhead{m-M}& 
\colhead{$\pi_{abs}$(mas)}
} 
\startdata
XZ Cyg&&&&&&\\
2&15.24&   K0V   &5.9&0.0&9.34$\pm$0.5&1.3$\pm$0.3\\
3&15&   G1.5V&4.6&0.3&10.39 0.5&1.0 0.2\\
4&12.55&   K2III &0.5&0.3&12.05 0.5&0.5 0.1\\
5&13.16&   F7V   &3.9&0.0&9.3 0.5&1.4 0.3\\
6&12.46&   G2V   &4.7&0.0&7.78 0.5&2.8 0.6\\
\\
UV Oct&&&&&&\\
9&15.92&G5V&5.1&0.1&10.8 0.5&0.7 0.2\\
10&15.35&K0V&5.9&0.4&9.5 0.5&1.6 0.4\\
11&14.91&G0V&4.2&0.2&10.7 0.5&0.8 0.2\\
12&13.12&K3III&0.3&0.3&12.8 0.5&0.3 0.1\\
13&14.45&F9V&4.2&0.2&10.2 0.5&1.0 0.2\\
\\
RZ Cep&&&&&&\\
17&14.9&   G1V   &4.5&1.4&10.36 0.5&1.6 0.4\\
18&16.3&   G1V   &4.5&2.1&11.76 0.5&1.2 0.3\\
19&16.16&   G2V   &4.7&1.3&11.48 0.5&0.9 0.2\\
20&15.31&   K0V   &5.9&1.1&9.41 0.5&2.2 0.5\\
21&14.06&   G1V   &4.5&1.1&9.52 0.5&2.1 0.5\\
22&12.52&   A1V   &0.9&1.7&11.61 0.5&1.1 0.2\\
\\
SU Dra&&&&&&\\
24&16.26&   G5V   &5.1&0.1&11.16 0.5&0.6 0.1\\
25&13.1&   F6V   &3.7&0.1&9.42 0.5&1.4 0.3\\
26&14.6&  K2.5V  &6.6&0.4&7.97 0.5&3.0 0.7\\
27&14.36&   G5V   &5.1&0.0&9.26 0.5&1.4 0.3\\
28&15.19&   F9V   &4.2&0.0&10.97 0.5&0.7 0.1\\
\\
\kp&&&&&&\\
31&12.65& K1.5III&0.6&0.1&12.1 0.5&0.4 0.1\\
32&14.24&G3V&4.8&0.2&9.4 0.5&1.4 0.3\\
33&14.55&   K1V   &6.2&0.1&8.4 0.5&2.2 0.5\\
34&15.94&   K1V   &6.2&0.2&9.8 0.5&1.2 0.3\\
35&14.84&   K0V   &5.9&0.0&8.9 0.5&1.6 0.4\\
36&15.31&   K0V   &5.9&0.1&9.4 0.5&1.4 0.3\\
37&15.7&K0V&5.9&0.1&9.8 1.0&1.2 0.5\\
\\
VY Pyx&&&&&&\\
39&12.45&K3III&0.3&0.3&12.15 0.5&0.4 0.1\\
40&15.31&F4V&3.3 &0.5&11.97 0.5&0.5 0.1\\
41&15.30&G8V&5.6 &0.5&9.72 0.5&1.4 0.3\\
42&14.41&F6V&3.7 &0.5&10.73 0.5&0.9 0.2\\
43&16.15&F4V&3.3&0.4&12.81 0.5&0.3 0.1\\
\enddata 
\end{deluxetable}

\begin{deluxetable}{llrr}
\tablewidth{0in}
\tablecaption{\kp~and Reference Star Relative Positions \tablenotemark{a} \label{tbl-POS}}
\tablehead{\colhead{FGS ID}&
\colhead{V} &
\colhead{$\xi$ \tablenotemark{b}} &
\colhead{$\eta$ \tablenotemark{b}} 
}
\startdata
\kp&4.23&-57.7608$\pm$0.0002&-156.2033$\pm$0.0002\\
31&12.65&-75.1709 0.0010&-47.2432 0.0012\\
32\tablenotemark{c}&14.23&0.0000 0.0003&0.0000 0.0003\\
33&14.53&-150.7073 0.0003&-195.2863 0.0003\\
34&15.81&48.7661 0.0007&-151.6338 0.0010\\
35&14.84&-70.5595 0.0004&-298.9709 0.0004\\
36&15.3&-149.3689 0.0003&-278.0559 0.0003\\
37&15.68&38.8649 0.0004&-59.3234 0.0005\\
\enddata
\tablenotetext{a}{epoch 2007.744}
\tablenotetext{b}{$\xi$ and $\eta$ are relative positions in arcseconds
}
\tablenotetext{c}{RA = 284\fdg230996, Dec = -67\fdg187251, J2000, epoch 2007.744}
\end{deluxetable}

\begin{center}
\begin{deluxetable}{l l l l l l}
\tablewidth{0in}
\tablecaption{\kp~and Reference Star Relative Proper Motion, Parallax and Space Velocity\label{tbl-PM}}
\tablehead{\colhead{ID}&
\colhead{$\mu_x$\tablenotemark{a}} &
\colhead{$\mu_y$\tablenotemark{a}} &
\colhead{$\pi_{abs}$\tablenotemark{b}} &
\colhead{V$_t$\tablenotemark{c}} 
 }
\startdata
\kp\tablenotemark{d}&-7.41$\pm$0.24&16.41$\pm$0.24&5.57$\pm$0.28&15.3$\pm$0.9\\
31&3.38 0.23&-4.79 0.22&1.82 0.25&15.2 2.5\\
32&29.47 0.73&-0.26 0.97&1.16 0.42&120.9 443.4\\
33&0.89 0.33&-0.28 0.35&0.17 0.27&25.2 51.5\\
34&-0.11 0.50&-15.51 0.44&0.26 0.42&280 1311\\
35&-8.80 1.15&-3.04 1.03&1.74 0.42&25.4 11.1\\
36&-9.79 0.39&-3.55 0.42&1.15 0.34&42.8 13.7\\
37&9.80 0.49&-3.31 0.44&1.46 0.33&33.5 8.9\\

\enddata
\tablenotetext{a}{$\mu_x$ and $\mu_y$ are relative motions along RA and Dec in mas
yr$^{-1}$ }
\tablenotetext{b}{Parallax in mas}
\tablenotetext{c}{V$_t = 4.74\times \mu/\pi_{abs}$}
\tablenotetext{d}{Modeled with equations 2 -- 5}
\end{deluxetable}
\end{center}



\begin{deluxetable}{r l l}
\tablecaption{CP2 Parallaxes, Proper Motions, and Absolute Magnitudes\label{tbl-SUM2}} 
\tablewidth{0pt}
\tablehead{
\colhead{Parameter} & \colhead{} &\colhead{}
}
\startdata
&\kp&VY Pyx\\
Duration (y)&2.23&2.63\\
Ref stars (\#)&7&5\\
Ref $\langle$V$\rangle$&14.75&14.72\\
Ref $\langle$B-V$\rangle$&0.91&0.79\\
\HST~$\mu$ (mas y$^{-1}$)&18.1$\pm$0.1&31.8$\pm$0.2\\
P.A. (\arcdeg)&335.5$\pm$0.1&20.5$\pm$0.1\\
V$_t$\tablenotemark{a} (\kms)&16.0$\pm$0.7&23.4$\pm$0.6\\
\HST~$\pi_{abs}$ (mas)&\kppi&\VYpi \\
Hip97 $\pi_{abs}$ (mas)&6.00$\pm$0.67&5.74$\pm$0.76\\
Hip07 $\pi_{abs}$ (mas)&6.52$\pm$0.77&5.01$\pm$0.44\\
LKH Corr&-0.02&-0.01\\
(m-M)$_0$&6.29&6.00\\
M$_V$&-1.99$\pm$0.11&+1.18$\pm$0.08\\
M$_K$&-3.52$\pm$0.11&-0.26$\pm$0.08\\
\enddata
\tablenotetext{a}{V$_t = 4.74\times \mu/\pi_{abs}$}
\end{deluxetable}

\begin{deluxetable}{r l l l l l}
\tablecaption{RRL Parallaxes, Proper Motions, and Absolute Magnitudes\label{tbl-SUM1}} 
\tablewidth{0pt}
\tablehead{
\colhead{Parameter} & \colhead{} &\colhead{}&\colhead{}&\colhead{}
}
\startdata
&XZ Cyg&UV Oct&RZ Cep&SU Dra&RR Lyr\\
Duration (y)&2.87&2.38&2.41&2.52&13.14\\
Ref stars (\#)&4&5&4&5&5\\
Ref $\langle$V$\rangle$&13.68&14.75&14.88&14.70&13.75\\
Ref $\langle$B-V$\rangle$&0.78&0.85&1.04&0.70&0.71\\
\HST~$\mu$ (mas y$^{-1}$)&86.1$\pm$0.1&133.4$\pm$0.2&214.4$\pm$0.3&90.7$\pm$0.2&222.5$\pm$0.1\\
P.A. (\arcdeg)&106.3$\pm$0.2&205.8$\pm$0.1&25.4$\pm$0.1&210.9$\pm$0.1&209.1$\pm$0.1\\
V$_t$\tablenotemark{a} (\kms)&245$\pm$22&368$\pm$20&400$\pm$27&307$\pm$25&333$\pm$13\\
\HST~$\pi_{abs}$ (mas)&\XZpi&\UVpi&\RZpi&\SUpi& \RRpi \\
Hip97 $\pi_{abs}$ (mas)&2.28$\pm$0.86&1.48$\pm$0.94&0.22$\pm$1.09&1.11$\pm$1.15&4.38$\pm$0.59\\
Hip07 $\pi_{abs}$ (mas)&2.29$\pm$0.84&2.44$\pm$0.81&0.59$\pm$1.48&0.20$\pm$1.13&3.46$\pm$0.64\\
LKH Corr&-0.09&-0.03&-0.05&-0.11&-0.02\\
(m-M)$_0$&8.99&8.85&8.02&9.38&7.13\\
M$_V$&+0.41$\pm$0.22&+0.35$\pm$0.13&+0.27$\pm$0.17&+0.40$\pm$0.25&+0.54$\pm$0.07\\
M$_K$&-0.29$\pm$0.22&-0.60$\pm$0.13&-0.40$\pm$0.16&-0.73$\pm$0.25&-0.65$\pm$0.07\\
\enddata
\tablenotetext{a}{Tangential velocity, V$_t = 4.74\times \mu/\pi_{abs}$}
\end{deluxetable}

\begin{deluxetable}{l l c c c c l}
\tablewidth{0in}
\tablecaption{K and V Zero-Points, a$_n$\label{tbl-KZP}}
\tablehead{\colhead{n} & \colhead{$\lambda$}&  \colhead{a(LKH)}&  \colhead{a(RP)}&\colhead{b\tablenotemark{1}}&\colhead{c\tablenotemark{2}}& \colhead{Notes}
}
\startdata
1&K$_s$&-0.56$\pm$0.02&-0.54$\pm$0.03&-2.38&0.08&\cite{Sol08}\\
2&K$_s$&-0.57~~0.03&-0.54~~0.03&-2.38& -&\cite{Sol06}, no [Fe/H]\\
3&K$_s$&-0.58~~0.04&-0.56~~0.04&-2.101&0.231&\cite{Bon03a}\\
4&K$_s$&-0.56~~0.02&-0.53~~0.04&-2.16& - &\cite{Dal04}\\
5&K$_s$&-0.57~~0.02&-0.53~~0.03&-2.71&0.12&\cite{Del06}\\
6&K$_s$&-0.56~~0.03&-0.54~~0.03&-2.11&0.05&\cite{Bor09}\\
\hline
7&V&0.45~~0.05&0.46~~0.03& - &0.214&\cite{Gra04}\\
\enddata
\tablenotetext{1}{b = log\,P coefficient}
\tablenotetext{2}{c = [Fe/H] coefficient}
\end{deluxetable}

\begin{deluxetable}{l l}
\tablewidth{0pt}
\tablecaption{RRL M$_V$ at [Fe/H]=-1.5\label{tbl-MvComp}} 
\tablehead{
\colhead{M$_V$} & \colhead{Source\tablenotemark{a}} 
}
\startdata
0.45$\pm$0.05&TP, this study, LKH\\
0.46 0.03&TP, this study, RP\\
0.40 0.22&TP, Koen \& Laney (1998)\nocite{Koe98} \\
0.61 0.16\tablenotemark{b}&TP, Benedict \etal (2002a,b), Feast (2002)\nocite{Fea02}\\
0.47 0.12&GC, Carretta \etal (2000)\nocite{Car00b}\\
0.62 0.11&HB, Carretta \etal (2000)\nocite{Car00b}\\
0.75 0.13&SP, Gould \& Popowski (1998)\nocite{Gou98}\\
0.55 0.12& SB, Cacciari \& Clementini (2003)\nocite{Cac03}\\
0.68 0.05&SP, Fernley \etal (1998a)\nocite{Fer98a}\\
\enddata
\tablenotetext{a}{SP = statistical parallax, GC = from subdwarf fits to globular clusters, HB = from trig parallax of field HB stars, SB = surface brightness, TP = trig parallax}
\tablenotetext{b}{Based on RR Lyrae only; includes an estimated cosmic dispersion component}
\end{deluxetable}

\begin{deluxetable}{l l l l l r l l l}
\tablewidth{0in}
\tablecaption{Globular Cluster Distance Moduli and Ages\label{tbl-DGC}}
\tablehead{\colhead{ID}&  \colhead{[Fe/H]\tablenotemark{a}}&  \colhead{E(B-V)}&\colhead{$\lambda$}&\colhead{m$_0$}& \colhead{M$_0$}& (m-M)$_0$& \colhead{Ref.\tablenotemark{b}} &\colhead{Age\tablenotemark{c}}
}
\startdata
M3&-1.57&0.01&V&15.62$\pm$0.05&0.45$\pm$0.11&15.17$\pm$0.12&1&\\
&&&K$_s$&13.93 0.04&-1.23\tablenotemark{d}&15.16 0.06&2&10.8$\pm$1.0\\
M4&-1.40&0.36&V&12.15 0.06&0.47 0.12&11.68 0.13&3&\\
&&&K$_s$&10.97 0.06&-0.52&11.48 0.08&4&$11.1^{ -1.4}_{+1.7}$\\
M15&-2.16&0.09&V&15.51 0.05&0.32 0.08&15.20 0.09&5&\\
&&&K$_s$&14.67 0.1&-0.52&15.18 0.11&4&12.1 1.0\\
M68&-2.08&0.04&V&15.51 0.01&0.33 0.08&15.18 0.08&6&\\
&&&K$_s$&14.35 0.04&-0.75&15.10 0.06&7&12.4 1.0\\
$\omega$ Cen&-1.84&0.11&V&14.2 0.02&0.38 0.09&13.82 0.09&8&\\
&&&K$_s$&13.05 0.06&-0.75&13.80 0.08&9& --\\
M92&-2.16&0.025&V&15.01 0.08&0.31 0.08&14.70 0.11&10&\\
&&&K$_s$&13.86 0.04&-0.78&14.64 0.06&11&13.1 1.1\\
\enddata
\tablenotetext{a}{ZW scale}
\tablenotetext{b}{1 \cite{Bnk06}; 2 \cite{But03}; 3 \cite{Cac79}; 4 \cite{Lon90}, $\langle$K$\rangle$ and error estimated from figure 1(c) at logP=-0.3; 5 \cite{Sil95}, table 6, RRL ab only; 6 \cite{Wal94}; 7 \cite{Dal06};  8 \cite{Ole03}; 9 \cite{Del06}, logP=-0.2; 10 \cite{Kop01}, table2, intensity averaged V; 11 \cite{Del05}, table 3, RRL ab only, $\langle$logP$\rangle$=-0.19.
}
\tablenotetext{c}{in Gy}
\tablenotetext{d}{M$_K$ errors, $\sigma = 0.05$ mag}
\end{deluxetable}

\begin{deluxetable}{l l l}
\tablewidth{0in}
\tablecaption{LMC Distance Moduli \label{tbl-DLMC}}
\tablehead{\colhead{Bandpass}& \colhead{(m-M)$_0$} &\colhead{Source}
}
\startdata
RRL & & \\
K$_s$&18.55$\pm$0.05 & 1\\
V&18.61 0.05&2\\
V&18.46 0.06& 3\\
\hline
\\
Reticulum Cluster RRL& & \\
K$_s$&18.50 0.03& 4\\
\hline
\\
Classical Cepheids & &\\
V& 18.52 0.06 &5\\
K&  18.48 0.04 & 5\\
W$_{VI}$&  18.51 0.04& 5\\
\enddata
\begin{flushleft}Notes:
\\ 1 LMC data from \cite{Bor09}\\
2 LMC data from \cite{Gra04}\\ 
3 LMC data from \cite{Sos03}\\
4 Reticulum cluster data from \cite{Dal04}\\
5 See \cite{Ben07}\\
The results for both the RRs and Cepheids are
from LKH corrected  absolute magnitudes.
\end{flushleft}
\end{deluxetable}
%
%

\clearpage

\clearpage
\clearpage

\clearpage

\clearpage

\clearpage

\clearpage

\end{document}